\documentclass[preprint,12pt]{elsarticle}
\usepackage{amssymb}
\usepackage{amsmath}
\usepackage{tabularx}
\usepackage{float}
\usepackage{graphicx}
\usepackage{adjustbox}
\usepackage{booktabs}
\usepackage{longtable}
\usepackage{subcaption}

\begin{document}

\begin{frontmatter}



\title{When Context Dominates: Multimodal Signatures of Takeover Readiness Under Varying Hazard and Cognitive Load Conditions}

\author[label1]{Shiva Azimi}

\author[label1]{Yasaman Hakiminejad} 

\author[label2]{Luis Gomero} 

\author[label3]{Elizabeth Pantesco} 

\author[label3]{Irene P. Kan} 

\author[label2]{Meltem Izzetoglu} 

\author[label1]{Arash Tavakoli} 

\affiliation[label1]{organization={Department of Civil and Environmental Engineering, Villanova University},
addressline={800 E Lancaster Ave},
city={Villanova},
postcode={19085},
state={PA},
country={US}}

\affiliation[label2]{organization={Department of Electrical and Computer Engineering, Villanova University},
addressline={800 E Lancaster Ave},
city={Villanova},
postcode={19085},
state={PA},
country={US}}

\affiliation[label3]{organization={Department of Psychological and Brain Sciences, Villanova University},
addressline={800 E Lancaster Ave},
city={Villanova},
postcode={19085},
state={PA},
country={US}}

\begin{abstract}
Semi-automated driving systems promise to reduce crashes by assisting with perception and control, yet they simultaneously introduce additional human factors challenges by requiring drivers to monitor automation and rapidly resume control when failures occur. Prolonged passive monitoring can degrade vigilance, delay reactions, and increase takeover risk, but the extent to which distraction, hazard context, and drivers’ underlying cognitive and physiological states jointly shape takeover performance remains insufficiently understood. This study investigates these interacting factors using a controlled, within-subjects driving simulator experiment that crosses two hazard types (dynamic pedestrian and static crash events) with three levels of secondary task engagement (no task, conversation, and working memory load). Driver responses were assessed using a multimodal sensing framework that integrates vehicle-dynamics measures, subjective workload ratings, autonomic physiology (electrodermal activity and heart rate variability), and prefrontal cortical activation measured with functional near-infrared spectroscopy. Results show that hazard context is the primary determinant of takeover behavior, with pedestrian events producing longer and more variable maneuvers and crash events yielding faster and more stable responses. Secondary tasks exerted smaller effects on objective vehicle control, while internal-state measures showed more variable task-related patterns. These findings highlight the importance of jointly considering environmental context and human state when evaluating takeover readiness and designing driver monitoring systems. This study lays the groundwork for adaptive, context-aware strategies that support safer human–automation collaboration in semi-automated vehicles.
\end{abstract}









\end{frontmatter}

\section{Introduction}

Motor vehicle crashes remain a leading cause of injury and death worldwide, with over 40,000 fatalities annually in the United States alone \cite{NHTSA_2024_Traffic_Fatality_Estimates}. Beyond their tragic human toll, these incidents impose substantial economic costs through lost productivity, property damage, and medical expenses. A large proportion of these crashes stem from human error, and among the most pervasive contributors is driver distraction \cite{kashevnik2021driver,pettitt2005defining}. Distraction, whether cognitive, visual, or manual, undermines situational awareness and delays reaction times during critical moments, making it one of the most pressing safety challenges in modern transportation systems \cite{regan2008driver,lee2014dynamics}.

Automated driving technologies have emerged as a powerful tool to address these challenges. By delegating aspects of perception, decision-making, and control to machines, automation promises to reduce the frequency of human-induced crashes, enhance roadway efficiency, and mitigate driver fatigue \cite{dunn2021investigating,zangi2022driver}. The progression from advanced driver-assistance systems (ADAS) to higher levels of automation has already begun to reshape the landscape of personal mobility. Automated features such as adaptive cruise control, lane-keeping assistance, and collision avoidance have demonstrated their capacity to support drivers in routine operations and reduce the cognitive burden associated with complex driving environments \cite{bosurgi2023effects,ariansyah2018analysis,jumaa2019advanced}.

However, the transition toward full automation is gradual, and most vehicles currently on the road operate in a semi-automated mode that still requires human oversight \cite{reagan2025behavior}. In these shared-control conditions, drivers must remain attentive and ready to take over when the system encounters limitations \cite{mueller2025finding}. Yet, evidence suggests that prolonged passive monitoring leads to reduced vigilance, slower takeover reactions, and increased crash risk when automation fails \cite{biondi2024vigilance}. In other words, the same automation designed to enhance safety can paradoxically introduce new forms of inattention and complacency. Understanding when and why drivers disengage—and how their cognitive and physiological states influence takeover performance—remains a key research priority \cite{lee2024understanding}.

Studies show that takeover performance is not determined by automation alone but emerges from a complex interplay between the driver’s internal state, the external driving scene, and the nature of distraction \cite{gold2013take,eriksson2017takeover}. For instance, cognitively demanding secondary tasks, emotionally charged events, or unexpected environmental stimuli can differentially affect how quickly and effectively a driver resumes control \cite{zeeb2016take,fridman2017autonomous,korber2016influence}. Investigating this interaction among distraction type, environmental context, and takeover behavior is essential for building next-generation driver monitoring systems that adapt to human variability. Such understanding can inform both vehicle interface design and real-time decision logic, ultimately leading to safer, more trustworthy human–autonomy collaboration. Despite these insights, empirical evidence linking distraction type, environmental context, and drivers’ underlying cognitive and physiological states to objective takeover performance remains limited.

To address this need, the present study systematically investigates how distraction, hazard context, and drivers’ cognitive and physiological states jointly shape takeover performance during semi-automated driving. Using a high-fidelity driving simulator, we implemented a controlled within-subjects experiment that crossed two representative hazard types (dynamic pedestrian and static crash events) with three levels of secondary task engagement (no task, conversational, and working-memory load). In addition to conventional vehicle-dynamics measures of takeover timing, control quality, and driver response strategies, we employed a multimodal sensing framework integrating autonomic physiology (heart rate variability (HRV) and electrodermal activity (EDA)), prefrontal cortical activation measured via functional near-infrared spectroscopy (fNIRS), and subjective workload assessments. This design enables not only the quantification of how drivers perform during takeovers, but also the identification of the cognitive and affective mechanisms underlying those behaviors. By linking observable driving actions with internal human states, this study lays the groundwork for adaptive driver monitoring systems, context-aware takeover strategies, and safer human–automation collaboration in semi-automated vehicles.

The remainder of this paper is organized as follows. Section~\ref{sec:background} reviews relevant background and related work. Section~\ref{sec:rq} presents the research questions and hypotheses. Section~\ref{sec:methodology} describes the experimental design, data collection framework, and feature extraction procedures. Section~\ref{sec:results} presents the results. Section~\ref{sec:discussion} discusses implications for driver monitoring and automated driving system design. Finally, Sections~\ref{sec:limitation} and~\ref{sec:conclusion} outline limitations, future work, and concluding remarks.

\section{Background Information} \label{sec:background}
\subsection{Automation and Shared Control in Driving}

Recent advances in vehicle automation have shifted the role of the human driver from a primary operator to a supervisor of automated systems. According to the Society of Automotive Engineers (SAE), Level~2 (partial automation) and Level~3 (conditional automation) represent a critical threshold in which control responsibilities are shared between the automated system and the human driver \cite{sae2018j3016}. In Level~3 automation, the system performs the dynamic driving task under specific operational conditions, but the human driver must remain prepared to intervene when the system issues a takeover request \cite{du2020predicting,kulkarni6073126preventing}. This shared-control paradigm introduces significant challenges, particularly regarding the timing and quality of transitions between automated and manual control.

The transition of control, commonly referred to as a \emph{takeover}, is not an instantaneous event, but rather a complex cognitive and physical process. Successful takeovers require drivers to regain situational awareness, interpret the system’s request, and physically re-engage with the vehicle controls \cite{hayashi2019driver}. During automated driving, drivers frequently engage in non-driving-related tasks (NDRTs), which may result in an ``out-of-the-loop'' (OOTL) state where situational awareness is reduced and response times increase \cite{du2020psychophysiological,roche2019behavioral}. Such degradation in awareness can significantly affect driver readiness during takeover events \cite{endsley2017toward,biondi2024vigilance}.

Several theoretical frameworks provide a foundation for understanding takeover behavior in semi-automated driving. Wickens’ Multiple Resource Theory posits that human performance is constrained by multiple, partially independent cognitive resources distributed across modalities such as visual, auditory, and cognitive processing, such that engagement in non-driving-related tasks (NDRTs) leads to competition for these limited resources and reduces the capacity available for monitoring the driving environment and responding to takeover requests \cite{roche2019behavioral}. Complementing this perspective, the concept of Situation Awareness (SA), originally introduced by Endsley \cite{ensley1995toward}, describes human decision-making in dynamic environments as a process of perception, comprehension, and projection of environmental elements. This framework has been widely applied in driving research to explain how drivers maintain awareness and respond to takeover and control transitions \cite{endsley2017here,kaber2016effect,heenan2014effects}. During automated driving, engagement in secondary tasks or prolonged passive monitoring can disrupt these processes, resulting in delayed hazard recognition and impaired decision-making. This degradation is closely related to the OOTL performance problem, whereby reduced engagement with the driving task leads to diminished awareness and slower, less effective responses when control must be resumed. In addition to these established frameworks, recent work by Liang et al.~\cite{liang2025predicting} highlights the role of perceived spare capacity, a self-reported construct reflecting drivers' perceived task demand and perceived capability of handling takeovers, demonstrating that lower perceived spare capacity is associated with longer takeover times (ToT). \textbf{These perspectives suggest that takeover performance is shaped by the interaction between task demands, cognitive resource allocation, and the driver’s level of situational awareness during automated driving.}

To address the limitations in drivers’ cognitive resources and situational awareness during takeover, a growing body of research has explored adaptive takeover request (TOR) strategies and driver assistance mechanisms to support drivers in semi-automated driving environments. For example, dynamic alerts that adapt to the driver's cognitive state have been shown to improve takeover performance in semi-automated driving environments \cite{umpaipant2024improving}. Such systems aim to bridge the gap between human and machine capabilities by providing context-aware assistance, including situational awareness support that guides the driver’s attention toward critical elements of the driving environment during takeover situations \cite{hayashi2019driver}. Similar approaches have also been explored through multimodal takeover alerts that combine visual, auditory, and haptic cues to enhance driver response \cite{yun2020multimodal}, as well as adaptive takeover timing strategies that adjust the initiation of takeover requests based on traffic context and driver readiness \cite{du2020predicting}.

\subsection{Factors Affecting Takeover Performance}

The literature identifies a wide range of factors that influence the latency and quality of driver takeovers. These factors are commonly categorized into driver-related, task-related, environmental, and system-related dimensions \cite{tan2025driver}.

\textbf{Driver-related factors:}
Individual driver characteristics such as age, driving experience, skill level, and psychological state influence takeover performance. While some studies suggest that static driver characteristics such as driving experience may have limited explanatory value for takeover reaction time \cite{du2020predicting}, others indicate that driver-state measures, including heart rate indices, galvanic skin response, gaze behavior, and perceived spare capacity, a self-reported construct from Task--Capability Interface (TCI) theory, are important predictors of the time required to regain control \cite{du2020predicting,liang2025predicting}. Related work has also highlighted the role of drivers’ cognitive state and workload in shaping takeover performance \cite{umpaipant2024improving}. Cognitive conditions such as fatigue and emotional state have also been identified as important factors. For example, prolonged exposure to conditionally automated driving can induce fatigue, which can be observed through physiological changes and may negatively affect takeover performance \cite{coyne2026effect}. Additionally, physiological-sensing approaches suggest that drivers’ affective and psychophysiological states, reflected through measures such as skin conductance, heart rate, mental workload, and engagement, provide useful information for estimating takeover readiness and safety-related outcomes, including reaction time, time-to-collision, and maximum acceleration \cite{deng2024predicting}.

\textbf{Task-related factors (NDRTs):}
The characteristics of non-driving-related tasks (NDRTs) represent one of the most widely studied influences on takeover performance. Engagement in NDRTs can significantly reduce driver situational awareness and readiness to intervene \cite{du2020predicting,shi2019explore}. Research has shown that the modality of NDRTs plays a critical role; for instance, visual tasks such as reading reduce the frequency of road-facing glances compared to auditory tasks, leading to slower and less stable takeover responses \cite{roche2019behavioral}. Increased cognitive load associated with secondary tasks may also be reflected in physiological changes, such as elevated heart rate or electrodermal activity, which have been linked to decreased takeover performance \cite{deng2024predicting,du2020psychophysiological}.

\textbf{Environmental factors:}
External driving conditions also play a role in takeover performance. Factors such as traffic density, roadway complexity, and weather conditions can increase the cognitive and physiological demands placed on the driver during takeover situations \cite{deng2024predicting,du2020psychophysiological}. For example, complex traffic environments may increase mental workload and physiological stress responses during control transitions \cite{deng2024predicting,du2020psychophysiological}. In addition to general environmental complexity, hazard characteristics, including the type of obstacle, its motion, and the urgency of the situation, can significantly influence drivers’ decision-making and response strategies during takeover events. Drivers’ responses also vary across overt and covert hazards; overt hazards often prompt earlier braking or slowing, and covert hazards have been associated with poorer timely hazard perception, higher vehicle speeds, greater braking deceleration, and higher collision rates. Hazard type has also been shown to influence initial response behavior, braking intensity, vehicle speed adjustments, and acceleration patterns during driver responses to hazards \cite{wei2022analysis,ventsislavova2016happens,tan2025driver,gold2013take}.

\textbf{System-related factors (TOR design):}
The design of takeover requests (TORs), including their timing, urgency, and modality, represents another important factor affecting takeover performance. Shorter lead times for TORs have been associated with increased physiological arousal and provide drivers with less time to regain situational awareness before resuming control \cite{du2020psychophysiological}. In addition, multimodal warnings that combine auditory, visual, and haptic cues generally result in faster and more reliable takeover responses compared to unimodal alerts \cite{kulkarni6073126preventing}. These findings highlight the importance of carefully designing human–machine interfaces to support safe transitions between automated and manual driving \cite{hayashi2019driver}.

\subsection{Driver State Estimation and Monitoring Approaches}

Ensuring safe transitions between automated and manual driving requires accurate estimation of the driver's readiness to regain control. Consequently, researchers have developed driver monitoring approaches that utilize behavioral observations, physiological signals, and vehicle-based indicators.

\textbf{Vehicle-based indicators:} Vehicle dynamics and control inputs provide another source of information for estimating driver engagement. Measures derived from steering behavior, pedal inputs, lane position, speed, and acceleration can reflect driver control quality and performance during takeover maneuvers. Studies have shown that vehicle-based data can also be used to detect driver distraction and environmental demand during driving \cite{kanaan2019using}. However, because these indicators typically capture behavior after control has been regained, they are often combined with physiological or behavioral sensing modalities to improve early detection of driver state and takeover readiness \cite{jirjees2025integrated}.

\textbf{Behavioral monitoring approaches:} Driver-facing cameras and eye-tracking systems are widely used to observe visible indicators of driver engagement. Measures such as gaze direction, glance behavior, head pose, and facial expressions are commonly used to estimate driver attention and situational awareness. Vision-based gaze tracking systems can also detect ``eyes-off-road'' behavior and other forms of visual distraction that indicate reduced driver readiness \cite{vicente2015driver}. In the context of automated driving, eye-tracking studies have shown that drivers' pre-takeover visual engagement, particularly attention directed toward the driving scene and relevant roadway elements, is associated with their situation awareness and preparedness to respond to takeover events \cite{liang2021using}. These behavioral monitoring approaches therefore provide important information about driver attentiveness and potential distraction during semi-automated driving.

\textbf{Physiological monitoring approaches:} Physiological sensing methods provide insight into the driver’s internal state by measuring autonomic responses associated with cognitive workload, stress, fatigue, and arousal. Signals such as heart rate, HRV, EDA, and respiration have been widely used to estimate drivers’ psychophysiological states during driving \cite{chowdhury2018sensor,al2024technologies}. These physiological markers can reveal changes in driver readiness even when observable behaviors remain stable. Recent studies have also shown that machine learning techniques applied to physiological signals can effectively detect driver stress, drowsiness, and fatigue, supporting their use in real-time driver monitoring systems \cite{siam2023automatic,sukumar2024physiological}.

\textbf{Neurophysiological monitoring approaches:} In addition to peripheral physiological signals, neurophysiological techniques such as electroencephalography (EEG) and fNIRS have been increasingly used to measure brain activity associated with attention, cognitive workload, and fatigue during driving tasks. EEG studies have shown that changes in brain activity can reflect variations in driver alertness and mental workload \cite{borghini2014measuring,lohani2019review}. These techniques allow researchers to directly observe cortical responses during driving and takeover scenarios, providing deeper insight into the neural mechanisms underlying driver readiness and cognitive processing \cite{wascher2023tracking}.

Recent studies have increasingly explored multimodal driver monitoring systems that integrate behavioral observations, physiological signals, and vehicle-based indicators. By combining multiple sensing modalities, these approaches aim to improve the robustness and accuracy of driver state estimation compared with single-sensor methods. Multimodal frameworks have therefore become an important direction in driver monitoring research for automated and semi-automated driving systems.

\subsection{Gaps}

Despite extensive research on driver takeover behavior in automated driving, several limitations remain in the current literature. Previous studies have examined factors influencing takeover performance, including takeover timing, driver readiness, and the effects of non-driving related tasks during automated driving \cite{eriksson2017takeover,zeeb2015determines}. In parallel, recent work has explored multimodal driver monitoring approaches that integrate behavioral, physiological, and vehicle-based signals to predict driver takeover behavior and readiness \cite{pakdamanian2021deeptake}.

However, \textbf{many existing studies examine these aspects in isolation.} Takeover research often focuses primarily on behavioral performance metrics such as reaction time or vehicle control actions, while driver monitoring research typically aims to detect driver states such as distraction or workload without directly linking these internal states to observable takeover behavior in specific driving situations. In addition, although secondary tasks are known to influence driver readiness and attention allocation, relatively limited work has examined how different types of cognitive distraction interact with hazard context to influence takeover responses. \textbf{Consequently, the relationship between hazard context, secondary task engagement, and drivers’ internal cognitive and physiological states in shaping takeover behavior remains largely unexplored.} To address this gap, the present study investigates how these factors jointly influence takeover performance in semi-automated driving using a multimodal sensing framework that integrates behavioral measures, physiological signals, and neurophysiological monitoring.

\section{Research Questions and Hypotheses}
\label{sec:rq}
To systematically examine how hazard context, distraction, and drivers’ internal cognitive and physiological states influence takeover behavior during semi-automated driving, this study addresses the following research questions:

\begin{itemize}
    \item \textbf{RQ1 (Hazard context):} How does hazard type, specifically dynamic pedestrian events versus static crash events, influence driver takeover performance?
    
    \item \textbf{RQ2 (Secondary task engagement):} How does increased cognitive load, induced through NDRTs, affect takeover behavior and perceived workload relative to a no-task baseline?
    
    \item \textbf{RQ3 (Interaction effects):} Do distraction and hazard context interact to jointly influence takeover performance?
    
    \item \textbf{RQ4 (Human-state mechanisms):} How are physiological and neural indicators of driver state associated with takeover performance and subjective workload?
\end{itemize}

Based on prior findings on vigilance decrement, cognitive load, and the OOTL performance problem in automated driving, we test the following hypotheses:

\begin{itemize}
    \item \textbf{H1:} Takeover performance will differ as a function of hazard type, with dynamic pedestrian hazards producing longer and more variable takeover maneuvers than static crash hazards.
    
    \item \textbf{H2:} Secondary cognitive tasks will degrade takeover performance relative to the no-task condition, resulting in slower or less stable control responses.
    
    \item \textbf{H3:} The negative effects of cognitive load on takeover performance will be amplified during dynamic pedestrian hazards, indicating an interaction between distraction and environmental complexity.
    
    \item \textbf{H4:} Secondary task engagement will increase perceived workload, particularly mental demand and effort, as measured by NASA-TLX.
    
    \item \textbf{H5:} Physiological and neural indicators of driver's psychophysiological state will vary systematically across task and hazard conditions, reflecting changes in cognitive demand.
    
    \item \textbf{H6:} Variations in physiological and neural measures will be associated with takeover outcomes, such that states indicative of reduced vigilance or elevated cognitive load will correspond to slower or less stable takeover performance.
\end{itemize}

\section{Methodology} \label{sec:methodology}

\subsection{Participant Recruitment and Demographics}

The study protocol was reviewed and approved by the Institutional Review Board (IRB) at Villanova University (IRB Protocol \#IRB-FY2024-59). A total of 38 licensed drivers participated in the experiment. Participants were recruited through university mailing lists, campus flyers, and word-of-mouth. Eligibility criteria required participants to be at least 18 years old and to hold a valid driver’s license. Volunteers could enroll either in the full study, which included both the longitudinal monitoring phase and the simulator experiment, or in a driving-session-only condition. A summary of participant demographic characteristics is provided in Table~\ref{tab:demo}.

\begin{table}[ht]
\centering
\small
\caption{Participant Demographics (N = 38)}
\label{tab:demo}
\begin{tabular}{p{0.45\linewidth} p{0.4\linewidth}}
\toprule
\textbf{Characteristic} & \textbf{Summary} \\
\midrule
Age (years) & $27.8 \pm 8.0$ \\
Driving experience (years) & $8.3 \pm 7.4$ \\
\midrule
Gender & \\
\hspace{1em}Male & 21 \\
\hspace{1em}Female & 16 \\
\hspace{1em}Other / Unspecified & 1 \\
\midrule
Ethnicity & \\
\hspace{1em}White & 25 \\
\hspace{1em}Asian & 5 \\
\hspace{1em}Black or African American & 3 \\
\hspace{1em}White, Other & 1 \\
\hspace{1em}White, Asian & 1 \\
\hspace{1em}Prefer not to say & 1 \\
\hspace{1em}Other / Unspecified & 2 \\
\bottomrule
\end{tabular}
\end{table}

\subsection{Study Design}

This study employed a within-subjects experimental design to investigate driver takeover performance during semi-automated driving under varying cognitive load and hazard conditions. The study started with a short manual driving session (2 minutes) to familiarize participants with the simulator environment and vehicle control dynamics. Following this familiarization phase, participants transitioned to the semi-automated driving environment, where they were exposed to a series of trials including various takeover scenarios. During each trial, participants initially traveled in automated driving mode, during which both lateral and longitudinal control were managed by the simulator. When the vehicle reached a predefined trigger location in the scenario, a takeover request was issued via an auditory prompt, instructing the participant to resume manual control in response to an upcoming hazard. The auditory message stated: ``Please take control of the vehicle and continue driving until I am able to resume.'' Participants were instructed to respond naturally and safely using braking, steering, or lane-changing maneuvers, as appropriate to the scenario.

The experimental design of the takeover scenarios followed a $2 \times 3$ factorial structure, crossing takeover hazard type with secondary task engagement. Two hazard types were implemented:

\begin{itemize}
    \item \textbf{Pedestrian hazard event}: a dynamic scenario in which a pedestrian unexpectedly entered the roadway, moved through the driving scene, and stopped in the road, requiring rapid perception and response.
    \item \textbf{Crash hazard event}: a static scenario involving a stopped or crashed vehicle positioned in the left lane of the roadway, creating a lane-obstruction hazard ahead.
\end{itemize}

\begin{figure}[htbp]
    \centering
    \includegraphics[width=0.45\textwidth]{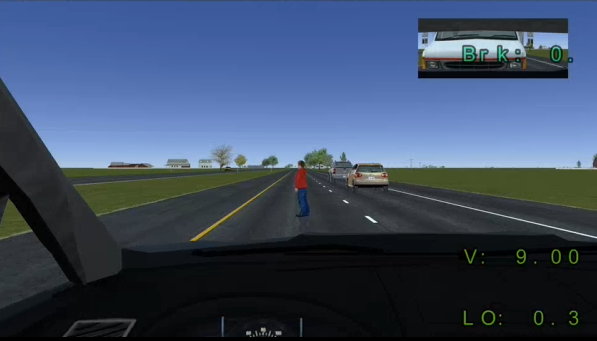}
    \hfill
    \includegraphics[width=0.45\textwidth]{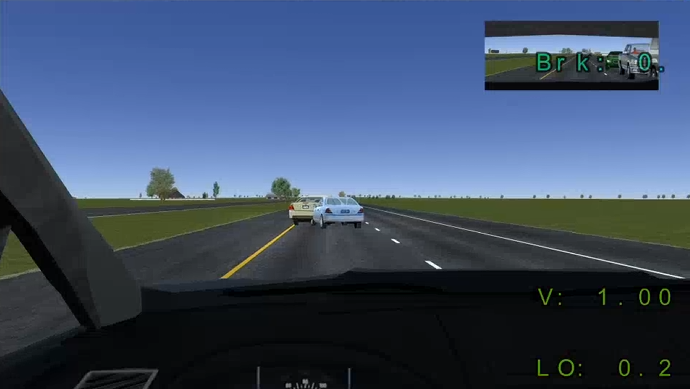}
    \caption{Representative snapshots of the two takeover hazard scenarios: (a) unexpected pedestrian event and (b) crash event.}
    \label{fig:hazard_scenarios}
\end{figure}

These hazard types were selected to represent qualitatively different takeover contexts, varying in perceived urgency, predictability, and motion dynamics. Figure~\ref{fig:hazard_scenarios} provides representative snapshots of the two hazard scenarios.

Secondary task engagement was manipulated across three levels to induce different degrees of cognitive load:

\begin{itemize}
    \item \textbf{No-task condition}, serving as a baseline with no additional cognitive demands.
    \item \textbf{N-Back task condition}, in which participants performed a 2-back auditory working memory task during automated driving.
    \item \textbf{Conversation task condition}, involving a naturalistic verbal interaction designed to simulate everyday in-vehicle conversations.
\end{itemize}


Each participant completed all combinations of hazard type and secondary task condition, resulting in six experimental conditions.

\subsection{Data Collection Framework}

\begin{figure*}[t]
\centering
\includegraphics[width=1\textwidth]{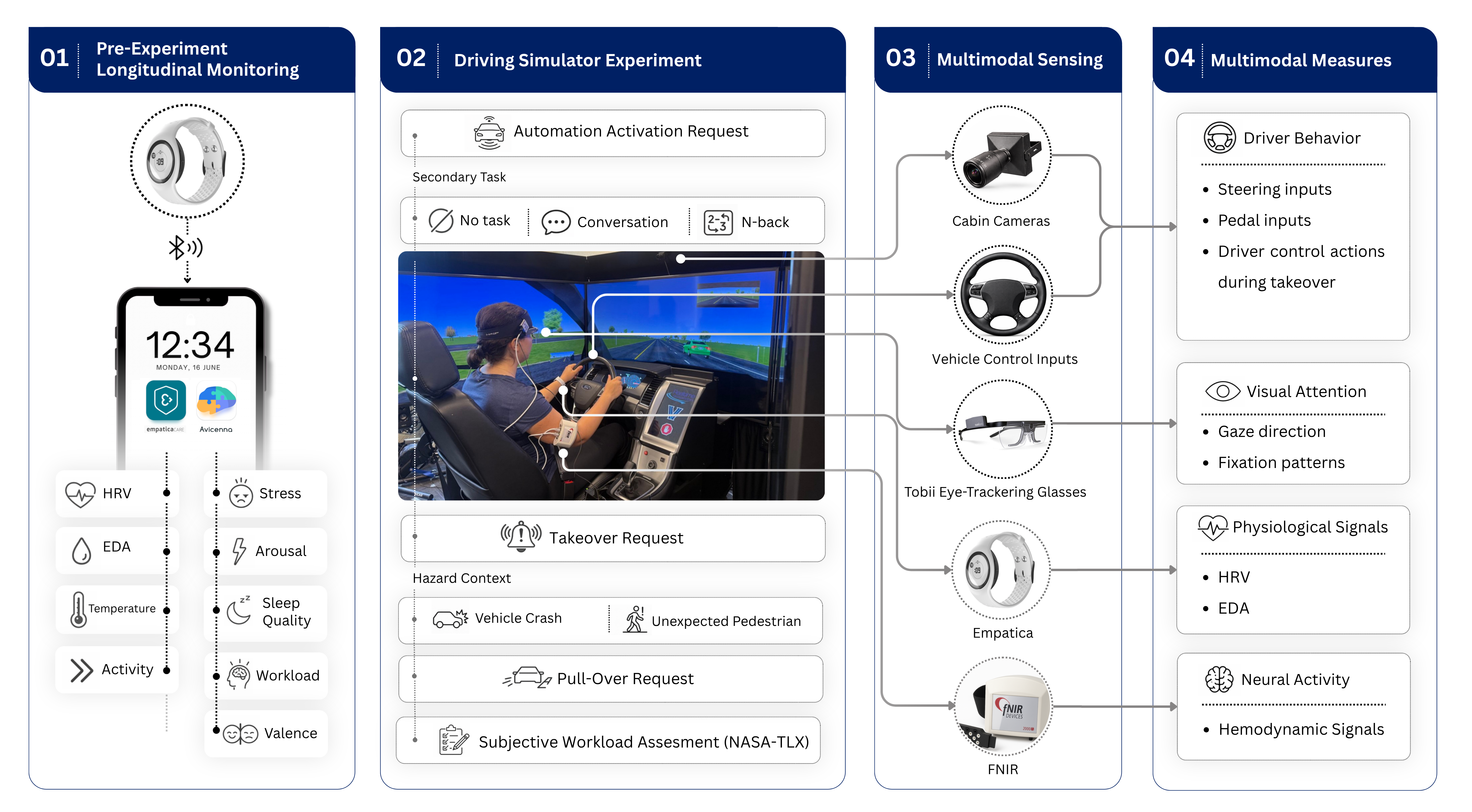}
\caption{Data collection framework: (1) pre-experiment longitudinal monitoring, (2) driving simulator experiment, (3) multimodal sensing, and (4) multimodal measures.}
\label{fig:workflow_framework}
\end{figure*}

Data were collected during a pre-experiment longitudinal monitoring phase and an in-lab driving simulation session, capturing behavioral, physiological, and neural responses. Figure~\ref{fig:workflow_framework} illustrates the overall workflow of the data collection framework, which integrates longitudinal monitoring with a controlled driving simulator experiment and multimodal sensing. The framework consists of four main components: (1) pre-experiment longitudinal monitoring (7-days) using wearable physiological sensing and experience sampling, (2) a driving simulator experiment in which participants experience takeover scenarios under varying hazard contexts and secondary task conditions, (3) multimodal sensing during the simulator session, including cabin cameras, vehicle control inputs, eye tracking, wearable physiological monitoring, and fNIRS to capture real-time driver responses, and (4) extraction of behavioral, visual attention, physiological, and neural measures used in the subsequent analyses. The present study is focused on the in-lab driving simulation session. 

\subsubsection{Pre-Experiment Longitudinal Monitoring}

Prior to the simulator session, a seven-day pre-experiment longitudinal monitoring phase was conducted. During this phase, participants wore a physiological sensing device, the Empatica EmbracePlus, on their wrists.  The device was worn continuously throughout daily activities, excluding periods of water exposure, and recorded heart rate, interbeat intervals (IBIs), EDA, skin temperature, and accelerometer-based motion data.

In parallel, participants completed daily experience sampling surveys delivered via a mobile application. These brief surveys assessed self-reported sleep quality, stress, emotional arousal, and subjective well-being. The combination of continuous physiological sensing and daily self-report measures provided a longitudinal profile of each participant’s recent psychophysiological state, which was later linked to in-simulator performance. More details on this portion of the study can be found in \cite{hakiminejad2026steering}.

\subsubsection{In-Lab Driving Simulation Session}

Following the longitudinal monitoring period, participants completed a controlled driving session in a high-fidelity driving simulator. During the simulation, participants were outfitted with multiple synchronized sensing modalities to capture real-time responses to takeover events, including a wearable smartwatch, an eye tracker, and an fNIRS device. 




\subsubsection{Self-Reported Data}

Following each takeover event, the vehicle was brought to a safe stop, after which participants completed a NASA Task Load Index (NASA-TLX) questionnaire to assess perceived workload associated with the preceding scenario. This procedure enabled trial-level subjective workload assessment aligned with behavioral and physiological measures.

Additionally, upon finishing the study, participant demographic information (e.g., age, gender, ethnicity, and driving experience) and self-reported measures regarding personality (Big Five Inventory), sleep quality, and driving behavior were collected. 

\subsubsection{Data Integrity and Synchronization}

To ensure data quality, sensor signals were monitored for completeness and integrity throughout data collection. Trials with missing or corrupted data in critical channels were flagged for exclusion during preprocessing. Because the study integrated multiple data streams, exclusion criteria were applied separately for each modality rather than using a single universal missing-data threshold. Data were excluded when missingness affected a critical portion of the corresponding modality and prevented reliable analysis of the planned study segment or event.

For physiological data, four participants were excluded because the device did not record during the relevant study segment. One participant was excluded because the driving simulator shut down due to a blackout during the study. For the fNIRS data, nine participants were excluded because of an incorrect data collection procedure that made the recordings unusable. For the Tobii eye-tracking data, two participants were excluded because the Tobii system stopped working during automation, resulting in missing data around the critical event window.

Thus, exclusions were based on modality-specific data usability and whether the missing or corrupted data prevented valid analysis of the corresponding behavioral, physiological, or neural measure.

All sensor streams were resampled or temporally aligned as necessary during post-processing to support joint analysis across behavioral, physiological, and neural domains.

This integrated data collection framework enabled comprehensive characterization of driver responses to takeover events, supporting subsequent feature extraction and statistical analysis across multiple levels of human performance.



\subsubsection{Vehicle-Based Sensing}

Vehicle dynamics data were collected from the high-fidelity driving simulator and recorded continuously throughout each trial at a sampling rate of 1 Hz. These signals included vehicle speed, longitudinal acceleration and deceleration, steering wheel angle, throttle position, brake input, lane position, and lane offset. These measurements provided continuous characterization of driver control behavior during automated driving and takeover events.

\subsubsection{Physiological Sensing}

Physiological data were collected using a wrist-worn Empatica EmbracePlus device. The device recorded multiple autonomic signals, including photoplethysmography (PPG) at 64 Hz for cardiac activity and interbeat intervals, EDA at 4 Hz for sympathetic arousal, peripheral skin temperature at 4 Hz, and three-axis accelerometer data at 32 Hz capturing motion and activity levels. These signals were recorded continuously during both the longitudinal monitoring phase and the in-lab simulation session.

\subsubsection{Neurophysiological Sensing}

Cortical brain activity was measured using a continuous-wave fNIRS system (fNIRS Imager 2000, Biopac Systems, Inc.). The system monitored prefrontal cortex (PFC) activity using light sources at wavelengths of 730, 805, and 850 nm. The focus on the PFC was motivated by both the device configuration and the established role of the PFC in higher-order cognitive processes relevant to simulator-based task performance, including attention, decision-making, and executive control. The sensor configuration consisted of 18 channels formed by 4 emitters and 12 detectors, including 16 long channels with a 25 mm inter-optode distance and 2 short channels with a 10 mm inter-optode distance. The full-forehead sensor pad provided bilateral coverage of the left and right hemispheres through a symmetrical arrangement across the forehead, enabling measurement of hemodynamic responses associated with neural activation. The 16 long channels were used for statistical analysis, as their larger source–detector separation allowed measurement of cortical hemodynamic activity. In contrast, the 2 short channels, which primarily capture superficial physiological signals and do not penetrate deeply enough to measure cortical activity, were used during preprocessing and quality assessment but were not included in the final statistical analyses or study conclusions. Data were sampled at 2 Hz during the simulator session.

\subsubsection{Eye-Tracking and Behavioral Sensing}

Visual attention and gaze behavior were recorded using Tobii Pro Glasses 2. The system captured binocular gaze data at 100 Hz, including gaze direction, fixation points, pupil size, and head motion through embedded inertial sensors. The eye tracker was calibrated individually for each participant prior to the experiment. Eye-tracking analyses and results are reported in a companion paper published prior to this analysis \citep{hakiminejad2026steering} and are not analyzed in the present study.

\subsubsection{Video-Based Behavioral Recording}

Driver behavior was additionally recorded using multiple in-cabin and simulator-mounted cameras. These cameras captured upper body posture, hand movements, and facial expressions throughout the experiment. These recordings were used for behavioral observation, task compliance verification, and contextual interpretation of driver actions.

All data streams were time-synchronized using system timestamps and aligned relative to experimental event markers (e.g., takeover request timing and task intervals).
\subsection{Feature Extraction}

To quantify driver responses during takeover events, multimodal features were extracted from vehicle dynamics, physiological signals, and neural measurements. Feature extraction focused on capturing both the timing and quality of driver control actions, as well as corresponding autonomic and cortical responses. All features were computed at the trial level and temporally aligned relative to the takeover request (TOR).

\subsubsection{Behavioral Features}

Behavioral features were extracted from vehicle dynamics signals to quantify driver takeover performance following the TOR \cite{cao2021towards,soares2021takeover}.

The takeover window was defined as the interval between the driver’s first meaningful control input following the takeover request and the moment at which the hazard was successfully cleared. Consistent with standard practice in takeover-performance research \cite{gold2013take}, the onset of manual control was identified when any control channel exceeded its minimum activation threshold. In this study, these thresholds were specifically defined as a brake pedal depression greater than 1\%, throttle input greater than 1\%, or steering wheel rotation exceeding 1°. This approach reliably distinguishes intentional driver actions from baseline sensor noise in the neutral position of the pedals and wheel.

Hazard clearance occurred when the driver either brought the vehicle to a complete stop before reaching the obstacle, performed a lane change, operationalized as crossing the lane centerline, prior to hazard contact, or bypassed the hazard by moving past its longitudinal position without stopping or changing lanes. Trials in which no identifiable control onset occurred or in which a valid clearance outcome could not be determined were excluded from analysis.

Within the resulting takeover window, a set of continuous behavioral metrics was computed. These included the mean, maximum, and standard deviation of speed, as well as the mean and variability of longitudinal acceleration. Negative acceleration values were used to quantify deceleration behavior, including the average deceleration, its variability, an impulse proxy derived from the time-integral of the magnitude of negative acceleration, and the number of discrete deceleration events, defined as periods where longitudinal acceleration dropped below –0.5 m/s² for at least 100 ms. The earliest control channel exceeding its threshold was classified as the driver’s first action type, and the temporal interval between this onset and the clearance moment was recorded as the takeover execution duration. These behavioral features served as the basis for subsequent statistical analyses comparing task conditions and event types.

All vehicle-dynamics signals were preprocessed to reduce noise and improve the stability of derivative-based measures. The velocity trace was first cleaned using a median filter with a window size of 5, which removes isolated spikes while preserving the underlying speed profile. Longitudinal acceleration was then derived from the filtered velocity signal using a central-difference method and subsequently smoothed with a moving-average window of size 5 to suppress high-frequency jitter. This two-step procedure is consistent with common practice for preparing simulator-based kinematic data in human factors and automated-driving research \cite{arien2024processing,montanino2013making,thiemann2008estimating}.

\subsubsection{Heart Rate Variability (HRV)}

Heart Rate Variability (HRV) features were derived from PPG signals as an index of autonomic nervous system activity during semi-automated driving \cite{shaffer2017overview}. HRV metrics were computed from interbeat intervals (IBIs) derived from systolic peak timestamps obtained from the Empatica EmbracePlus device \cite{embraceplus}.


Raw systolic peak timestamps were recorded in Coordinated Universal Time (UTC) and converted to local time (America/New York) to ensure alignment with simulator event markers. For each participant, all available systolic peak files from the final recording day were aggregated. Timestamps were converted to epoch seconds and seconds-of-day to facilitate synchronization with the simulator timeline.

Simulation start and end times were identified from experiment marker files and used to define the valid analysis window. When available, simulation boundaries were anchored to the nearest detected systolic peaks to improve temporal alignment between physiological data and simulator events. Participants for whom valid simulation boundaries could not be identified, or for whom no systolic peaks were present within the simulation window, were excluded from further HRV analysis. As a result, three participants were removed from the HRV analyses due to missing or invalid simulation boundaries or absence of usable systolic peak data during the simulation period.

Root Mean Square of Successive Differences (RMSSD) was computed for experimental task conditions corresponding to No Task, Conversation, and N-Back, each occurring twice per participant during automated driving. Event start and end times were obtained from the experiment marker file and anchored to the nearest systolic peaks within the simulation window.

Each event interval was subdivided into three temporal segments: the first 5 s following event onset, the last 5 s preceding event offset, and a middle segment defined as the remaining portion of the event after excluding the first and last 5 s. The middle segment was only defined for events with a total duration exceeding 10 s. Only the middle segment was used for HRV analysis, as it was intended to capture sustained autonomic activity during the task while minimizing transient effects related to event onset and offset.


Given the cognitive nature of the tasks examined in this study, RMSSD 
was selected as the primary cardiac metric, as it specifically indexes 
short-term parasympathetic modulation and has been established as the 
recommended time-domain measure of vagal tone in psychophysiological 
research \cite{laborde2017heart}, with demonstrated sensitivity to 
cognitive stress and workload \cite{cipryan2016within,kim2018stress}. RMSSD was computed from successive artifact-filtered IBIs as:

\[
\mathrm{RMSSD}
=
\sqrt{
\frac{1}{N - 1}
\sum_{i=1}^{N-1}
\left( IBI_{i+1} - IBI_i \right)^2
}
\]

where $IBI_i$ denotes the duration of the $i$-th interbeat interval in milliseconds, and $N$ denotes the total number of IBIs within the analyzed segment \cite{laborde2017heart}.

For each middle segment, systolic peak times were first sorted, and interbeat intervals (IBIs) were computed as the differences between consecutive peaks. Segments containing fewer than four systolic peaks were excluded. To reduce the influence of motion artifacts and implausible beat detections, segments were also excluded if any successive change in IBIs exceeded 2.5 s, indicating a likely artifact or missed beat. RMSSD was computed only for segments meeting these criteria.

All HRV measures were computed at the trial level and retained along with metadata describing participant identity, task condition, event timing, and the number of systolic peaks contributing to each estimate.

\subsubsection{Electrodermal Activity (EDA)}

EDA features were computed from skin conductance signals as an indicator of sympathetic nervous system activation associated with cognitive and emotional arousal. Raw EDA signals were visually inspected to identify signal dropouts and motion-related artifacts. Identified artifacts were mitigated using standard filtering and interpolation procedures.


EDA data from the Empatica device were processed at the task-event level. For each participant, available EDA files were loaded, non-finite values were removed, timestamps were standardized, and samples were sorted chronologically. Timestamps were converted from UTC to local time and restricted to the simulation interval defined by the start and end markers used in other data modalities. The simulation-level EDA signal was then exported to Ledalab, where continuous decomposition analysis (CDA) with optimization level 4 was used to separate the signal into tonic and phasic components. Decomposition quality was evaluated by correlating the original EDA signal with the reconstructed signal obtained from the sum of the tonic and phasic components.

Task-level EDA features were extracted for automated-driving events defined in the marker file. For each event, EDA samples falling between the event start and end times were selected. Mean tonic activity and mean phasic activity were calculated from the Ledalab-derived components. Phasic skin conductance responses were identified as peaks in the phasic component using a minimum peak prominence threshold of 0.02. The number of SCRs, SCR rate per minute, and phasic area under the curve were then computed for each event, with AUC estimated using trapezoidal integration and adjusted for the expected 4 Hz sampling rate.

\subsubsection{Functional Near-Infrared spectroscopy (fNIRS)}

Functional Near-Infrared spectroscopy (fNIRS) data were extracted to quantify task-related hemodynamic changes in prefrontal cortex during takeover events.

Raw optical intensity signals at 730 and 850 nm were first inspected for channels showing abnormally low magnitudes, indicative of poor optode coupling, calibration errors, or insufficient photon counts, and for saturated signals, identified as static plateaus beyond the device’s predefined threshold. To assess signal quality and detect noise contamination, the Scalp Coupling Index (SCI) was computed. Specifically, optical density (OD) signals were derived from the raw intensities and band-pass filtered with a 5th-order Butterworth filter (0.8–2.3 Hz) to isolate the cardiac pulsation component. The SCI was then calculated as the correlation between the two heartbeat-derived components across wavelengths. Channels with SCI \(<\) 0.70 were excluded due to insufficient scalp coupling. Additionally, to identify potential crosstalk or ambient light interference, the raw light intensity measured when both wavelengths were off was correlated with the intensity data for the 730 nm and 850 nm wavelengths, on a per-channel basis. Channels exhibiting a high correlation (r \(\ge\) 0.7) between the ambient light and either wavelength were flagged as corrupted, as such strong correlations indicate broadband contamination that prevents wavelength-based differentiation of HbO and HbR, rendering those channels unsuitable for further analysis.

Anatomical and physiological variations across participants, such as differences in forehead curvature, tissue thickness, or hair interference, also affected signal quality, particularly in regions where optodes could not maintain good contact. The SCI provided an objective criterion for identifying and removing such low-quality channels.

Motion artifacts were corrected using targeted spline interpolation in segments showing abrupt deviations exceeding 2.5 times the normalized standard deviation within the cardiac frequency band.

Subsequently, the modified Beer–Lambert Law (mBLL) was used to convert the corrected intensity data into relative concentration changes of oxyhemoglobin (HbO) and deoxyhemoglobin (HbR) with respect to a 5-s baseline. Differential pathlength factors (DPFs) were estimated as a function of wavelength and participant age, and the following molar extinction coefficients were used for each wavelength: 

\begin{itemize}
    \item $eHbR_{730} = 1.1022$ 
    \item $eHbO_{730} = 0.390$
    \item $eHbR_{850} = 0.69132$
    \item $eHbO_{850} = 1.058$
\end{itemize}

To minimize physiological confounds, short-separation channel signals (HbO and HbR) were used as regressors in an ordinary least squares (OLS) regression to remove extracerebral components from the long-separation channels (n = 16 maximum per participant, depending on SCI) \cite{izzetoglu2020short, izzetoglu2020effects}. The resulting corrected long-channel signals were then processed with a Daubechies wavelet filter (order 2, type 5) to reduce motion-related noise. Next, a high-pass filter (Window-based FIR filter, 0.009 Hz cutoff) was applied to remove slow drifts, followed by a low-pass filter (Window-based FIR filter, 0.08 Hz cutoff) to eliminate physiological artifacts, including heart rate, respiration rate, and Mayer waves.

For analysis, HbO and HbR data were epoched separately for each experimental condition (no-task, conversational task, and n-back task) and for each event type (dynamic pedestrian scenario or static road-crash scenario), yielding six distinct conditions. Each epoch was baseline-corrected by subtracting the mean of the first 5 s of its respective baseline period. Finally, mean HbO and HbR concentration changes relative to baseline were computed for each valid channel and condition, and these values were used for subsequent statistical comparisons.

\subsection{Statistical Analysis}

To account for the hierarchical structure of the data and the repeated observations contributed by the same participant, statistical analyses were conducted using linear mixed-effects models (LMMs). LMMs extend standard linear regression by incorporating both fixed effects, which represent population-level influences, and random effects, which capture variability associated with grouping factors such as participants. This makes LMMs particularly well suited for repeated-measures designs, as they explicitly model the non-independence of observations within individuals \cite{finucane2007translational,schober2021linear}. In contrast to traditional repeated-measures approaches, LMMs provide greater flexibility in handling unbalanced data, unequal numbers of observations across participants, and partially missing measurements without requiring exclusion of complete cases \cite{little1999summary}.

Following \cite{bates2015fitting}, a linear mixed-effects model can be expressed as:

\begin{equation}
Y_{ij} = \beta_0 + \sum_{k} \beta_k X_{k,ij} + b_i + \epsilon_{ij}
\end{equation}

where $Y_{ij}$ denotes the response variable for observation $j$ of participant $i$, $\beta_0$ is the fixed intercept, $\beta_k$ are fixed-effect coefficients associated with predictors $X_{k,ij}$, $b_i$ is the random effect associated with participant $i$ (assumed $b_i \sim \mathcal{N}(0, \sigma_b^2)$), and $\epsilon_{ij}$ is the residual error term (assumed $\epsilon_{ij} \sim \mathcal{N}(0, \sigma^2)$). This formulation allows the model to capture both systematic effects of experimental factors and subject-specific deviations from the overall mean.

In the present study, $Y_{ij}$ denotes outcome measures related to takeover performance, physiological responses, and subjective workload, whereas $X_{k,ij}$ represents experimental conditions including hazard type and secondary task.

Separate LMMs were fitted for each dependent variable of interest. Experimental conditions were included as fixed effects, comprising hazard type, secondary task, and their interaction, and participant was modeled as a random intercept to account for repeated measurements within individuals. Interaction terms were included to assess whether the effect of one factor depended on the level of another. This modeling approach enabled estimation of population-level effects while accounting for within-subject dependence.

Statistical analyses were conducted in R using the \texttt{lme4} package \cite{bates2015package,brown2021introduction}. Model fitting was performed using maximum likelihood estimation. Model assumptions were evaluated through visual inspection of residual distributions and residual-versus-fitted plots to assess approximate normality and homoscedasticity. Statistical significance was evaluated at an $\alpha$ level of 0.05. When significant main or interaction effects were observed, follow-up pairwise comparisons were conducted with appropriate adjustments for multiple comparisons.

For categorical behavioral outcomes, including first action type and final action type, multinomial logistic mixed-effects models were additionally employed. These models account for nominal outcomes with more than two categories while incorporating both fixed effects of experimental conditions and random effects for participants.

Following standard formulations \cite{agresti2013categorical}, the model can be expressed as:

\begin{equation}
\log \left( \frac{P(Y_{ij} = c)}{P(Y_{ij} = c_0)} \right)
= \beta_{0c} + \sum_{k} \beta_{kc} X_{k,ij} + b_{i}
\end{equation}

where $P(Y_{ij} = c)$ denotes the probability that observation $j$ of participant $i$ belongs to category $c$, and $c_0$ represents the reference category. The coefficients $\beta_{0c}$ and $\beta_{kc}$ are category-specific fixed effects, and $b_i$ is a participant-level random intercept accounting for repeated measurements.

Models were implemented in R using the \texttt{brms} package \cite{burkner2017brms}, which fits Bayesian multilevel models via Hamiltonian Monte Carlo sampling. Posterior estimates were summarized using means and 95\% credible intervals.

\section{Results} \label{sec:results}

\subsection{Behavioral Features}

The following results summarize the effects of task condition, event type, and their interaction on each takeover metric as estimated by the linear mixed-effects models. 

\begin{figure}[htbp]
\centering
\includegraphics[width=0.8\linewidth]{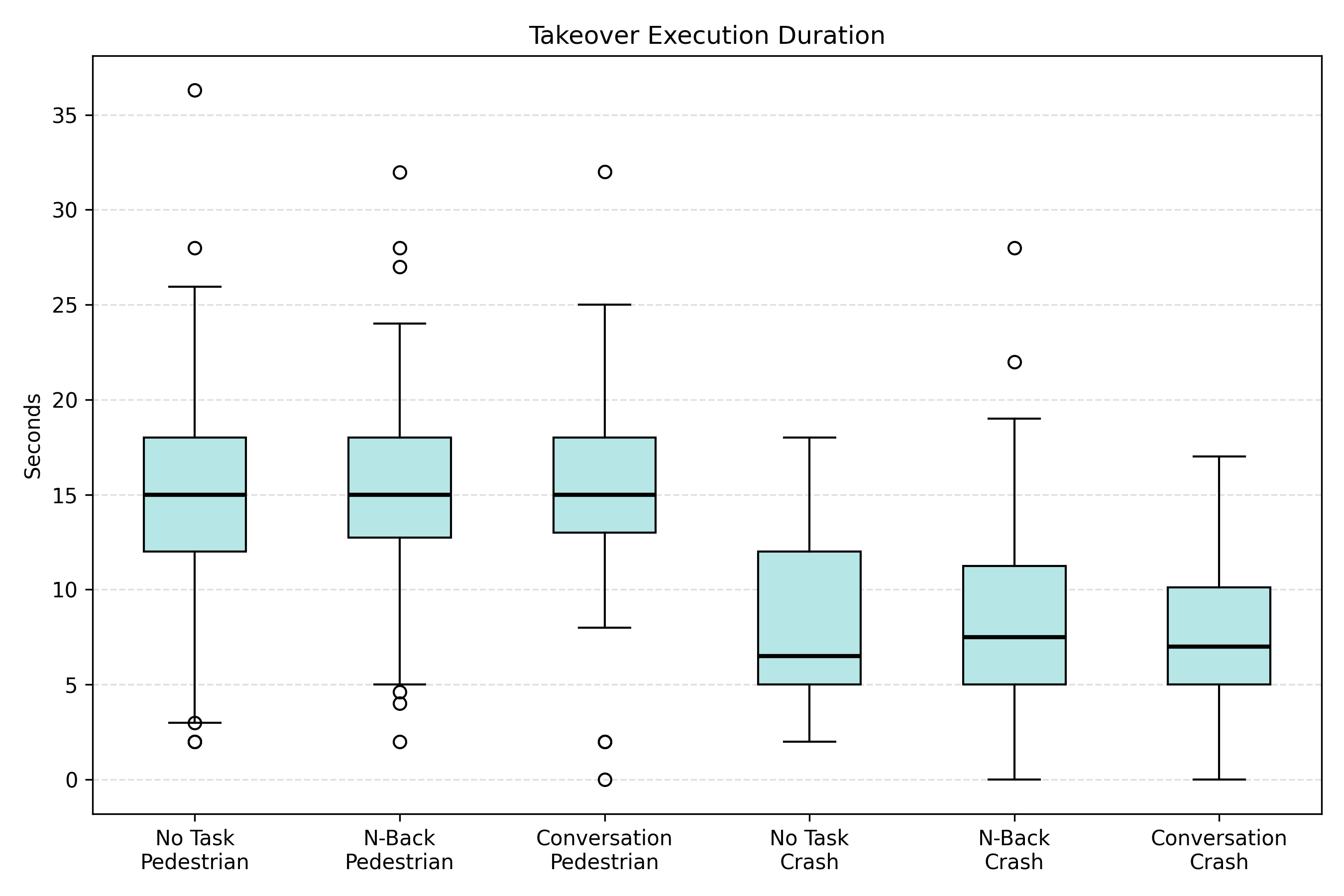}
\caption{Distribution of takeover execution duration across conditions.}
\label{fig:box_exec_duration}
\end{figure}

Figure~\ref{fig:box_exec_duration} illustrates the distribution of takeover
execution duration across conditions. Execution durations were significantly shorter for Crash hazards than for Unexpected Pedestrian hazards.

Table~\ref{tab:lme_exec_duration} shows the corresponding mixed-effects model estimates. Takeover execution duration was significantly shorter during Crash hazards than during Unexpected Pedestrian hazards ($\beta = -6.46$, $t(176.24) = -5.30$, $p < 0.001$), indicating faster completion of the takeover maneuver in the static hazard context. Secondary task engagement did not significantly affect takeover execution duration relative to the No Task reference (Conversation: $p = 0.722$; N-Back: $p = 0.965$), and the event-related difference was consistent across task conditions.

\begin{table}[H]
\centering
\caption{Linear mixed-effects model results for takeover execution duration.
Reference levels are \emph{No Task} and \emph{Unexpected Pedestrian}.}
\label{tab:lme_exec_duration}

\small 
\setlength{\tabcolsep}{4pt} 

\begin{adjustbox}{max width=\linewidth} 
\begin{tabular}{lccccccc}
\hline
Term & Coef & StdErr & DF & $t$ & $p$ & 95\% CI$_{lower}$ & 95\% CI$_{Upper}$ \\
\hline
Intercept & 15.0529 & 0.9662 & 172.49 & 15.58 & $<$ .001 & 13.1457 & 16.9600 \\
Conversation & $-$0.4311 & 1.2102 & 175.86 & $-$0.36 & 0.722 & $-$2.8194 & 1.9572 \\
N-Back & 0.0542 & 1.2195 & 176.24 & 0.04 & 0.965 & $-$2.3526 & 2.4609 \\
Crash & $-$6.4647 & 1.2195 & 176.24 & $-$5.30 & $<$ .001 & $-$8.8715 & $-$4.0580 \\
Conversation $\times$ Crash & $-$0.5070 & 1.7247 & 176.26 & $-$0.29 & 0.769 & $-$3.9107 & 2.8967 \\
N-Back $\times$ Crash & 0.3810 & 1.7313 & 176.44 & 0.22 & 0.826 & $-$3.0357 & 3.7976 \\
\hline
\end{tabular}
\end{adjustbox}

\end{table}

\begin{figure}[htbp]
\centering
\includegraphics[width=0.8\linewidth]{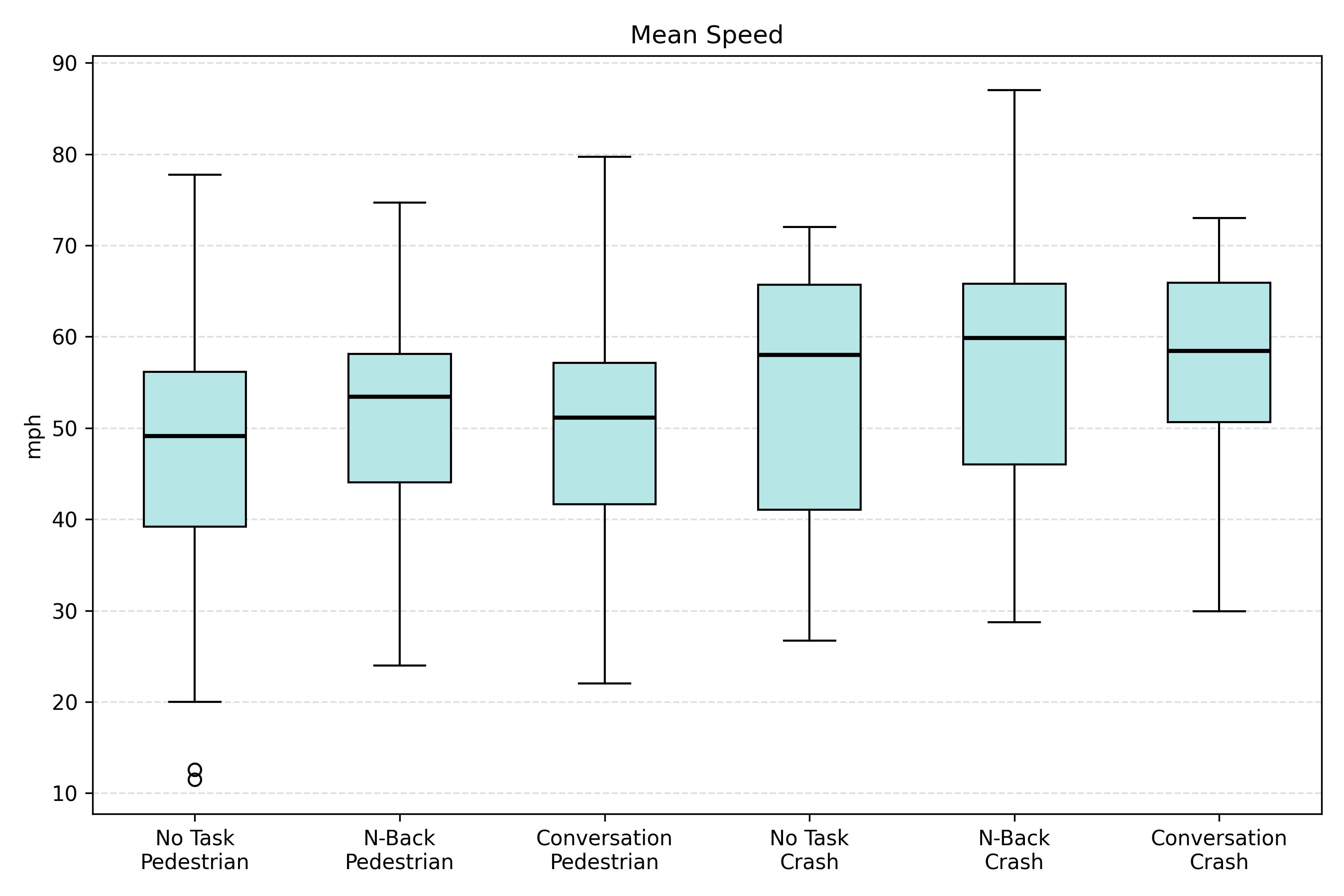}
\caption{Distribution of mean speed during the takeover window across conditions.}
\label{fig:box_speed_mean}
\end{figure}

Figure~\ref{fig:box_speed_mean} illustrates the distribution of mean speed
during the takeover window across conditions. Mean speed appears higher during Crash hazards than during Unexpected Pedestrian hazards.

Table~\ref{tab:lme_speed_mean} shows the corresponding mixed-effects model estimates. Mean speed during the takeover window was significantly higher during Crash hazards than during Unexpected Pedestrian hazards ($\beta = 5.82$, $t(172.46) = 2.20$, $p = 0.029$). Secondary task engagement did not significantly affect mean speed relative to the No Task reference (Conversation: $p = 0.429$; N-Back: $p = 0.172$), and the event type difference was similar across task conditions.

\begin{table}[htbp]
\centering
\caption{Linear mixed-effects model results for mean speed.
Reference levels are No Task and Unexpected Pedestrian.}
\label{tab:lme_speed_mean}

\small
\setlength{\tabcolsep}{4pt}

\begin{adjustbox}{max width=\linewidth}
\begin{tabular}{lccccccc}
\hline
Term & Coef & StdErr & DF & $t$ & $p$ & 95\% CI$_{lower}$ & 95\% CI$_{upper}$ \\
\hline
Intercept & 47.6499 & 2.1364 & 160.84 & 22.30 & 4.37e$-$51 & 43.4308 & 51.8690 \\
Conversation & 2.0956 & 2.6449 & 172.46 & 0.79 & 0.4293 & $-$3.1250 & 7.3162 \\
N-Back & 3.6293 & 2.6449 & 172.46 & 1.37 & 0.1718 & $-$1.5913 & 8.8499 \\
Crash & 5.8163 & 2.6449 & 172.46 & 2.20 & 0.0292 & 0.5957 & 11.0369 \\
Conversation $\times$ Crash & 2.1232 & 3.7703 & 172.84 & 0.56 & 0.5741 & $-$5.3185 & 9.5649 \\
N-Back $\times$ Crash & $-$1.2409 & 3.7703 & 172.84 & $-$0.33 & 0.7425 & $-$8.6825 & 6.2008 \\
\hline
\end{tabular}
\end{adjustbox}

\end{table}

\begin{figure}[htbp]
\centering
\includegraphics[width=0.8\linewidth]{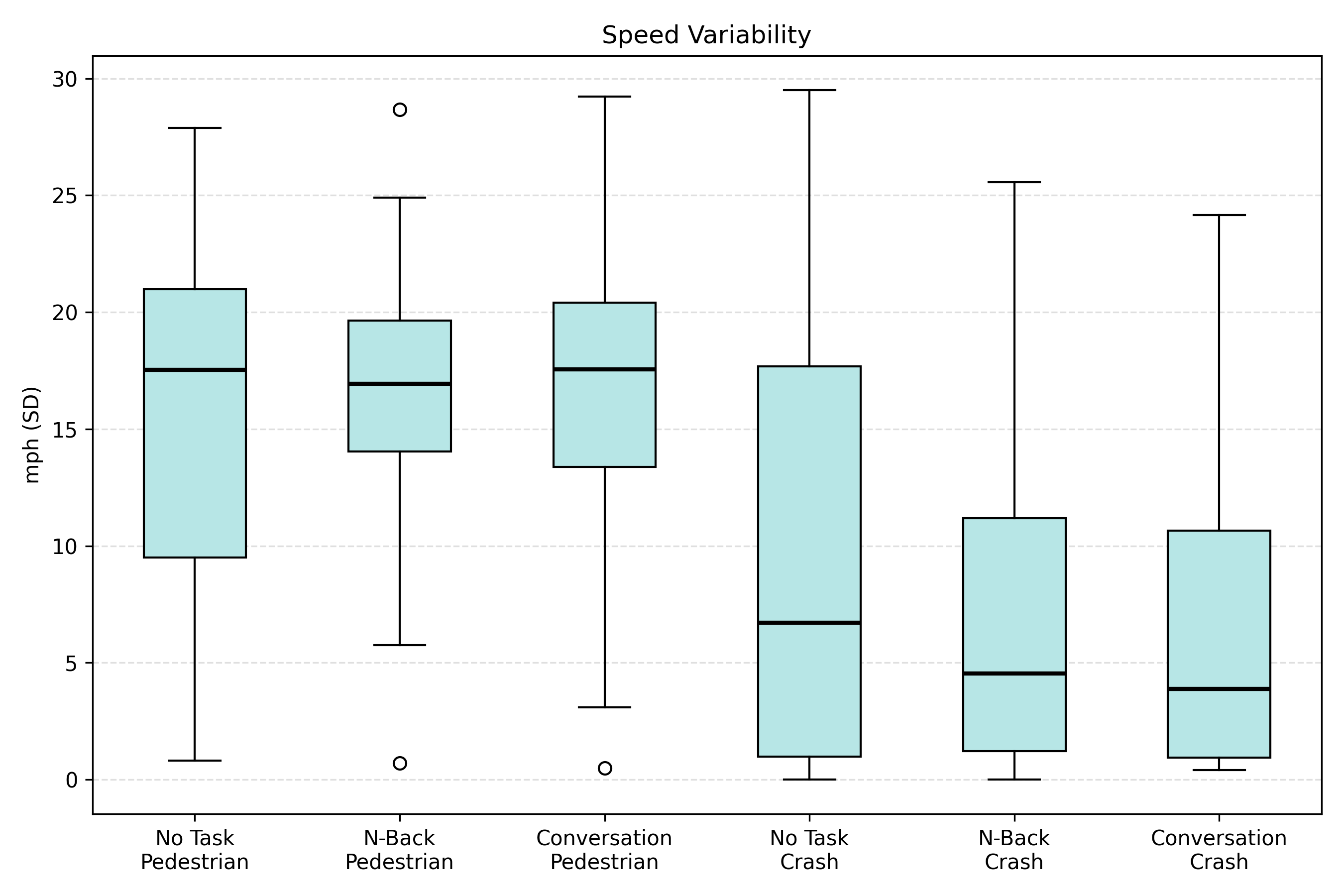}
\caption{Distribution of speed variability during the takeover window across conditions.}
\label{fig:box_speed_sd}
\end{figure}

Figure~\ref{fig:box_speed_sd} illustrates the distribution of speed variability
during the takeover window across conditions. Speed variability appears lower during Crash hazards than during Unexpected Pedestrian hazards, indicating more stable speed control in the static hazard context.

Table~\ref{tab:lme_speed_sd} shows the corresponding mixed-effects model estimates. Speed variability during the takeover window was significantly lower during Crash hazards than during Unexpected Pedestrian hazards ($\beta = -5.46$, $t(173.19) = -3.61$, $p < 0.001$). Task condition did not significantly affect speed variability (ps $> .05$); however, conversational engagement altered the pattern of speed regulation across hazard types, with a stronger reduction in speed variability during Crash events ($\beta = -4.93$, $t(173.57) = -2.28$, $p = 0.024$).

\begin{table}[htbp]
\centering
\caption{Linear mixed-effects model results for speed variability (speed SD).
Reference levels are No Task and Unexpected Pedestrian.}
\label{tab:lme_speed_sd}

\small
\setlength{\tabcolsep}{4pt}

\begin{adjustbox}{max width=\linewidth}
\begin{tabular}{lccccccc}
\hline
Term & Coef & StdErr & DF & $t$ & $p$ & 95\% CI$_{lower}$ & 95\% CI$_{upper}$ \\
\hline
Intercept & 15.3297 & 1.2158 & 164.24 & 12.61 & 6.31e$-$26 & 12.9291 & 17.7303 \\
Conversation & 1.4921 & 1.5136 & 173.19 & 0.99 & 0.3256 & $-$1.4953 & 4.4796 \\
N-Back & 1.1441 & 1.5136 & 173.19 & 0.76 & 0.4508 & $-$1.8434 & 4.1315 \\
Crash & $-$5.4567 & 1.5136 & 173.19 & $-$3.61 & 0.0004 & $-$8.4442 & $-$2.4693 \\
Conversation $\times$ Crash & $-$4.9292 & 2.1575 & 173.57 & $-$2.28 & 0.0235 & $-$9.1875 & $-$0.6708 \\
N-Back $\times$ Crash & $-$3.2085 & 2.1575 & 173.57 & $-$1.49 & 0.1388 & $-$7.4668 & 1.0498 \\
\hline
\end{tabular}
\end{adjustbox}

\end{table}

\begin{figure}[htbp]
\centering
\includegraphics[width=0.8\linewidth]{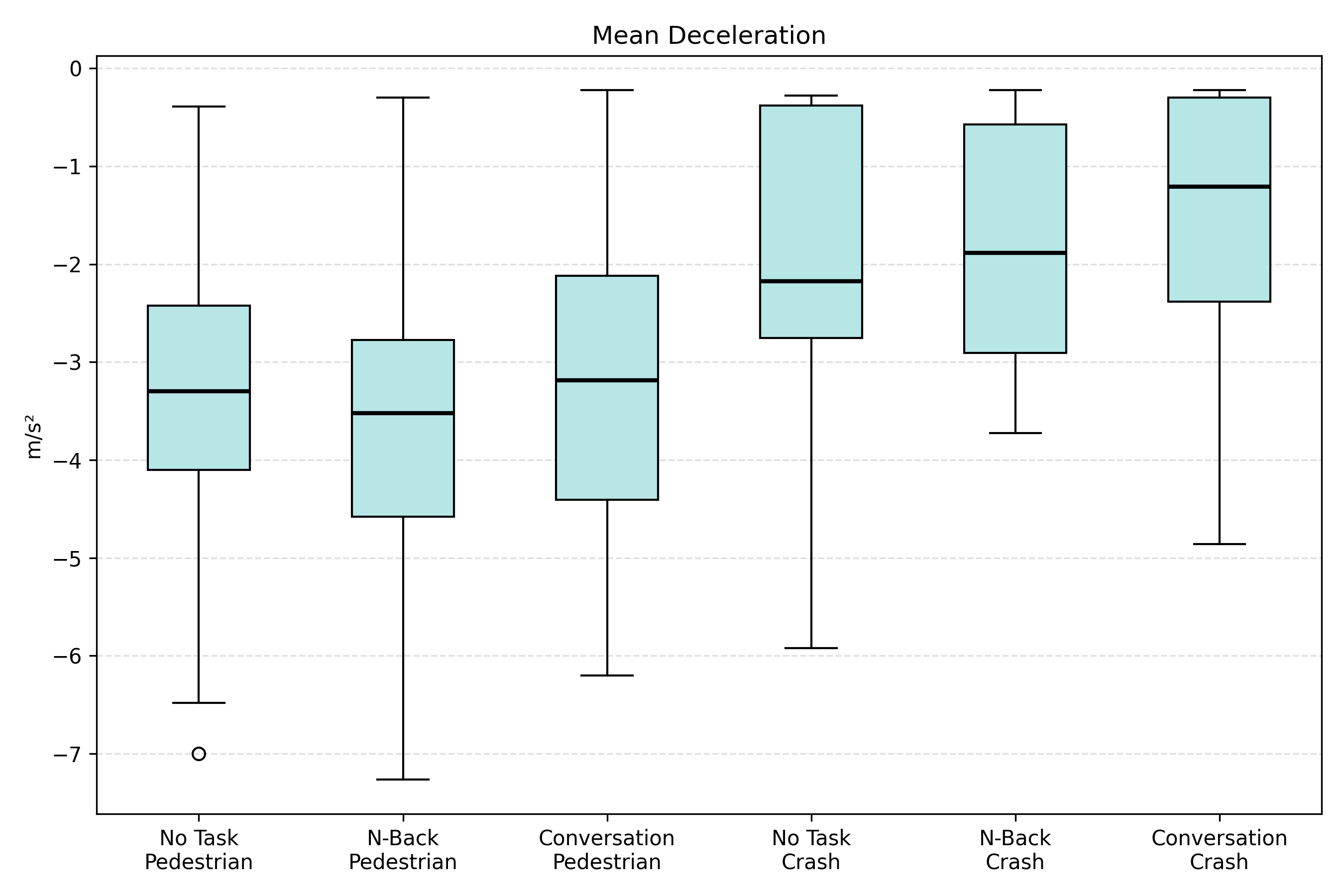}
\caption{Distribution of mean deceleration during the takeover window across conditions.}
\label{fig:box_decel_mean}
\end{figure}

Figure~\ref{fig:box_decel_mean} illustrates the distribution of mean deceleration during the takeover window across conditions. Mean deceleration appears less negative during Crash hazards than during Unexpected Pedestrian hazards, indicating lower average braking intensity in the static hazard context.

Table~\ref{tab:lme_decel_mean} shows the corresponding mixed-effects model estimates. Mean deceleration during the takeover window differed significantly by event type, with Crash hazards associated with less negative mean deceleration than Unexpected Pedestrian hazards ($\beta = 1.20$, $t(156.41) = 3.86$, $p < 0.001$). Task condition did not significantly affect mean deceleration (ps $> .05$), and the event-related difference was consistent across task conditions.

\begin{table}[H]
\centering
\caption{Linear mixed-effects model results for mean deceleration (m/s$^{2}$).
Reference levels are No Task and Unexpected Pedestrian.}
\label{tab:lme_decel_mean}

\small
\setlength{\tabcolsep}{4pt}

\begin{adjustbox}{max width=\linewidth}
\begin{tabular}{lccccccc}
\hline
Term & Coef & StdErr & DF & $t$ & $p$ & 95\% CI$_{lower}$ & 95\% CI$_{upper}$ \\
\hline
Intercept & $-$3.2740 & 0.2401 & 153.11 & $-$13.64 & 3.19e$-$28 & $-$3.7483 & $-$2.7996 \\
Conversation & $-$0.0468 & 0.2977 & 153.81 & $-$0.16 & 0.8753 & $-$0.6349 & 0.5413 \\
N-Back & $-$0.3065 & 0.3002 & 154.67 & $-$1.02 & 0.3089 & $-$0.8996 & 0.2866 \\
Crash & 1.1997 & 0.3112 & 156.41 & 3.86 & 0.00017 & 0.5850 & 1.8143 \\
Conversation $\times$ Crash & 0.4422 & 0.4444 & 156.45 & 1.00 & 0.3213 & $-$0.4356 & 1.3200 \\
N-Back $\times$ Crash & 0.5075 & 0.4461 & 156.83 & 1.14 & 0.2570 & $-$0.3736 & 1.3886 \\
\hline
\end{tabular}
\end{adjustbox}

\end{table}

\begin{figure}[htbp]
\centering
\includegraphics[width=0.8\linewidth]{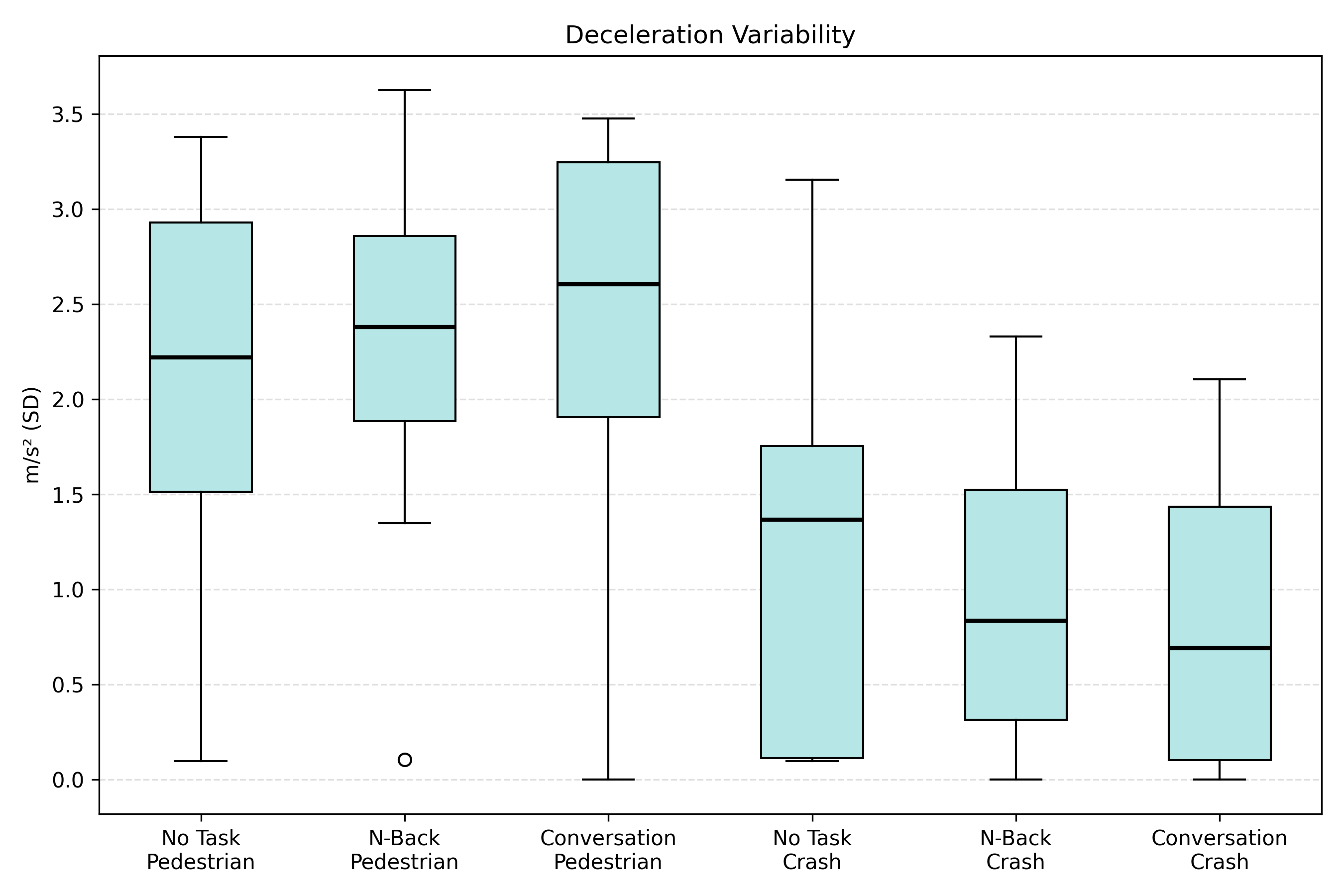}
\caption{Distribution of deceleration variability during the takeover window across conditions.}
\label{fig:box_decel_sd}
\end{figure}

Figure~\ref{fig:box_decel_sd} illustrates the distribution of deceleration variability during the takeover window across conditions. Deceleration variability appears lower during Crash hazards than during Unexpected Pedestrian hazards, indicating more consistent braking in the static hazard context.

Table~\ref{tab:lme_decel_sd} shows the corresponding mixed-effects model estimates. Deceleration variability, quantified as the standard deviation of deceleration (m/s$^{2}$), differed significantly by event type, with Crash hazards associated with lower variability than Unexpected Pedestrian hazards ($\beta = -0.94$, $t(159.67) = -4.93$, $p < 0.001$). Task condition did not significantly affect deceleration variability (ps $> 0.05$); however, the event-related reduction in variability differed under Conversation compared with the No Task reference ($\beta = -0.63$, $t(160.03) = -2.32$, $p = 0.022$).

\begin{table}[H]
\centering
\caption{Linear mixed-effects model results for deceleration variability (deceleration SD; m/s$^{2}$).
Reference levels are No Task and Unexpected Pedestrian.}
\label{tab:lme_decel_sd}

\small
\setlength{\tabcolsep}{4pt}

\begin{adjustbox}{max width=\linewidth}
\begin{tabular}{lccccccc}
\hline
Term & Coef & StdErr & DF & $t$ & $p$ & 95\% CI$_{lower}$ & 95\% CI$_{upper}$ \\
\hline
Intercept & 2.1455 & 0.1364 & 182.78 & 15.73 & 8.12e$-$36 & 1.8764 & 2.4145 \\
Conversation & 0.3090 & 0.1833 & 156.30 & 1.69 & 0.0939 & $-$0.0531 & 0.6711 \\
N-Back & 0.2001 & 0.1848 & 157.17 & 1.08 & 0.2803 & $-$0.1648 & 0.5651 \\
Crash & $-$0.9423 & 0.1911 & 159.67 & $-$4.93 & 2.04e$-$06 & $-$1.3198 & $-$0.5649 \\
Conversation $\times$ Crash & $-$0.6333 & 0.2729 & 160.03 & $-$2.32 & 0.0216 & $-$1.1723 & $-$0.0942 \\
N-Back $\times$ Crash & $-$0.4246 & 0.2739 & 160.40 & $-$1.55 & 0.1230 & $-$0.9655 & 0.1163 \\
\hline
\end{tabular}
\end{adjustbox}

\end{table}

\begin{figure}[htbp]
\centering
\includegraphics[width=0.8\linewidth]{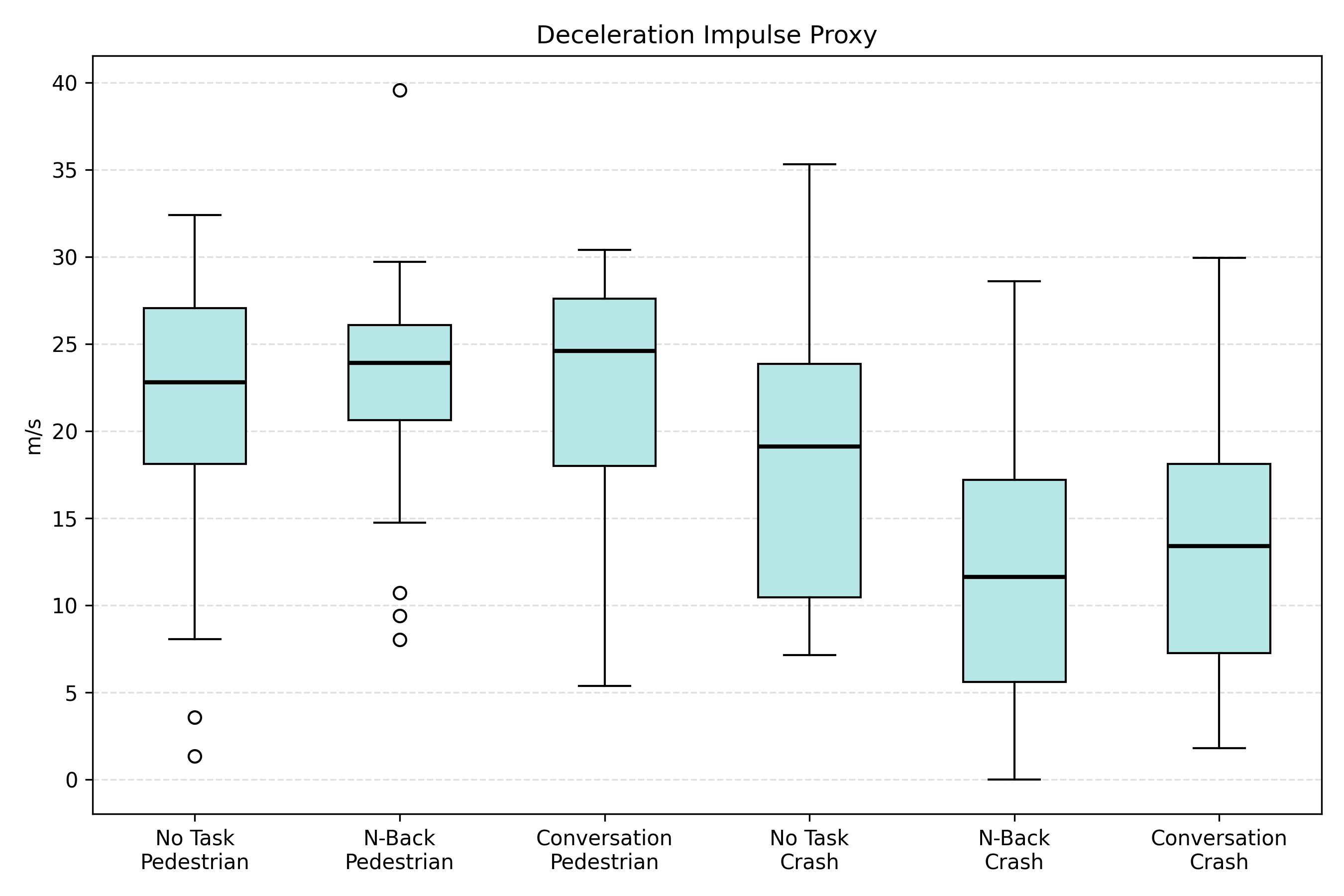}
\caption{Distribution of deceleration
impulse during the takeover window across conditions.}
\label{fig:box_decel_impulse}
\end{figure}

Figure~\ref{fig:box_decel_impulse} illustrates the distribution of deceleration
impulse during the takeover window across conditions. Deceleration impulse appears similar across hazard types, suggesting no clear difference in cumulative braking effort.

Table~\ref{tab:lme_decel_impulse} shows the corresponding mixed-effects model estimates. Deceleration impulse proxy (m/s), reflecting cumulative braking effort during the takeover window, did not show a significant main effect of event type ($\beta = -3.15$, $t(129.02) = -1.63$, $p = 0.106$). However, significant task-by-event effects were observed, with both Conversation ($\beta = -6.55$, $t(126.72) = -2.32$, $p = 0.022$) and N-Back ($\beta = -7.55$, $t(124.61) = -2.81$, $p = 0.006$) associated with a larger reduction in deceleration impulse during Crash hazards relative to Unexpected Pedestrian hazards.

\begin{table}[htbp]
\centering
\caption{Linear mixed-effects model results for deceleration impulse proxy (m/s).
Reference levels are No Task and Unexpected Pedestrian.}
\label{tab:lme_decel_impulse}

\small
\setlength{\tabcolsep}{4pt}

\begin{adjustbox}{max width=\linewidth}
\begin{tabular}{lccccccc}
\hline
Term & Coef & StdErr & DF & $t$ & $p$ & 95\% CI$_{lower}$ & 95\% CI$_{upper}$ \\
\hline
Intercept & 20.9847 & 1.2702 & 142.93 & 16.52 & 5.83e$-$35 & 18.4739 & 23.4955 \\
Conversation & 1.3386 & 1.6327 & 120.57 & 0.82 & 0.4139 & $-$1.8940 & 4.5712 \\
N-Back & 1.7485 & 1.6429 & 119.87 & 1.06 & 0.2893 & $-$1.5042 & 5.0013 \\
Crash & $-$3.1490 & 1.9344 & 129.02 & $-$1.63 & 0.1060 & $-$6.9762 & 0.6783 \\
Conversation $\times$ Crash & $-$6.5464 & 2.8169 & 126.72 & $-$2.32 & 0.0217 & $-$12.1208 & $-$0.9721 \\
N-Back $\times$ Crash & $-$7.5540 & 2.6889 & 124.61 & $-$2.81 & 0.0058 & $-$12.8758 & $-$2.2322 \\
\hline
\end{tabular}
\end{adjustbox}

\end{table}

\begin{figure}[H]
\centering
\includegraphics[width=0.8\linewidth]{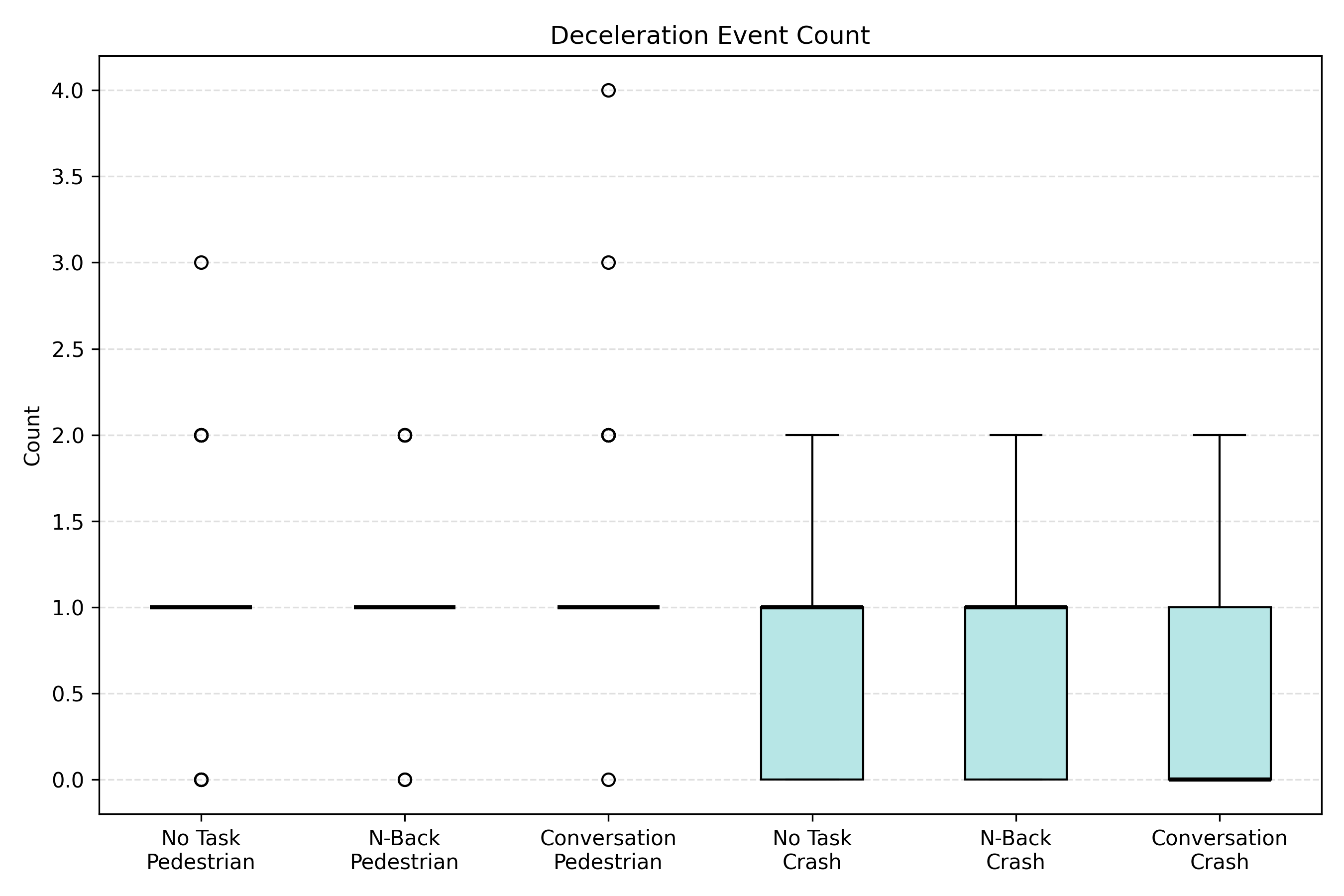}
\caption{Distribution of deceleration
event count during the takeover window across conditions.}
\label{fig:box_decel_count}
\end{figure}

Figure~\ref{fig:box_decel_count} illustrates the distribution of deceleration
event count during the takeover window across conditions. Deceleration event count appears lower during Crash hazards than during Unexpected Pedestrian hazards, indicating fewer distinct braking actions in the static hazard context.

Table~\ref{tab:lme_decel_count} shows the corresponding mixed-effects model estimates. Deceleration event count (number of discrete braking events) differed significantly by event type, with Crash hazards associated with fewer deceleration events than Unexpected Pedestrian hazards ($\beta = -0.50$, $t(173.93) = -3.92$, $p < 0.001$). Task condition did not significantly affect deceleration event count (ps $> .05$), and the event-related difference was similar across task conditions.

\begin{table}[htbp]
\centering
\caption{Linear mixed-effects model results for deceleration event count (events).
Reference levels are No Task and Unexpected Pedestrian.}
\label{tab:lme_decel_count}

\small
\setlength{\tabcolsep}{4pt}

\begin{adjustbox}{max width=\linewidth}
\begin{tabular}{lccccccc}
\hline
Term & Coef & StdErr & DF & $t$ & $p$ & 95\% CI$_{lower}$ & 95\% CI$_{upper}$ \\
\hline
Intercept & 1.1081 & 0.1027 & 161.25 & 10.79 & 8.75e$-$21 & 0.9054 & 1.3108 \\
Conversation & 0.1415 & 0.1267 & 173.93 & 1.12 & 0.2656 & $-$0.1086 & 0.3916 \\
N-Back & $-$0.0048 & 0.1267 & 173.93 & $-$0.04 & 0.9695 & $-$0.2549 & 0.2452 \\
Crash & $-$0.4961 & 0.1267 & 173.93 & $-$3.92 & 0.00013 & $-$0.7462 & $-$0.2461 \\
Conversation $\times$ Crash & $-$0.2129 & 0.1806 & 174.30 & $-$1.18 & 0.2400 & $-$0.5694 & 0.1435 \\
N-Back $\times$ Crash & 0.1421 & 0.1806 & 174.30 & 0.79 & 0.4326 & $-$0.2144 & 0.4985 \\
\hline
\end{tabular}
\end{adjustbox}

\end{table}

\subsection{Subjective Workload Measure (NASA-TLX)}

\begin{figure}[htbp]
\centering
\includegraphics[width=\linewidth]{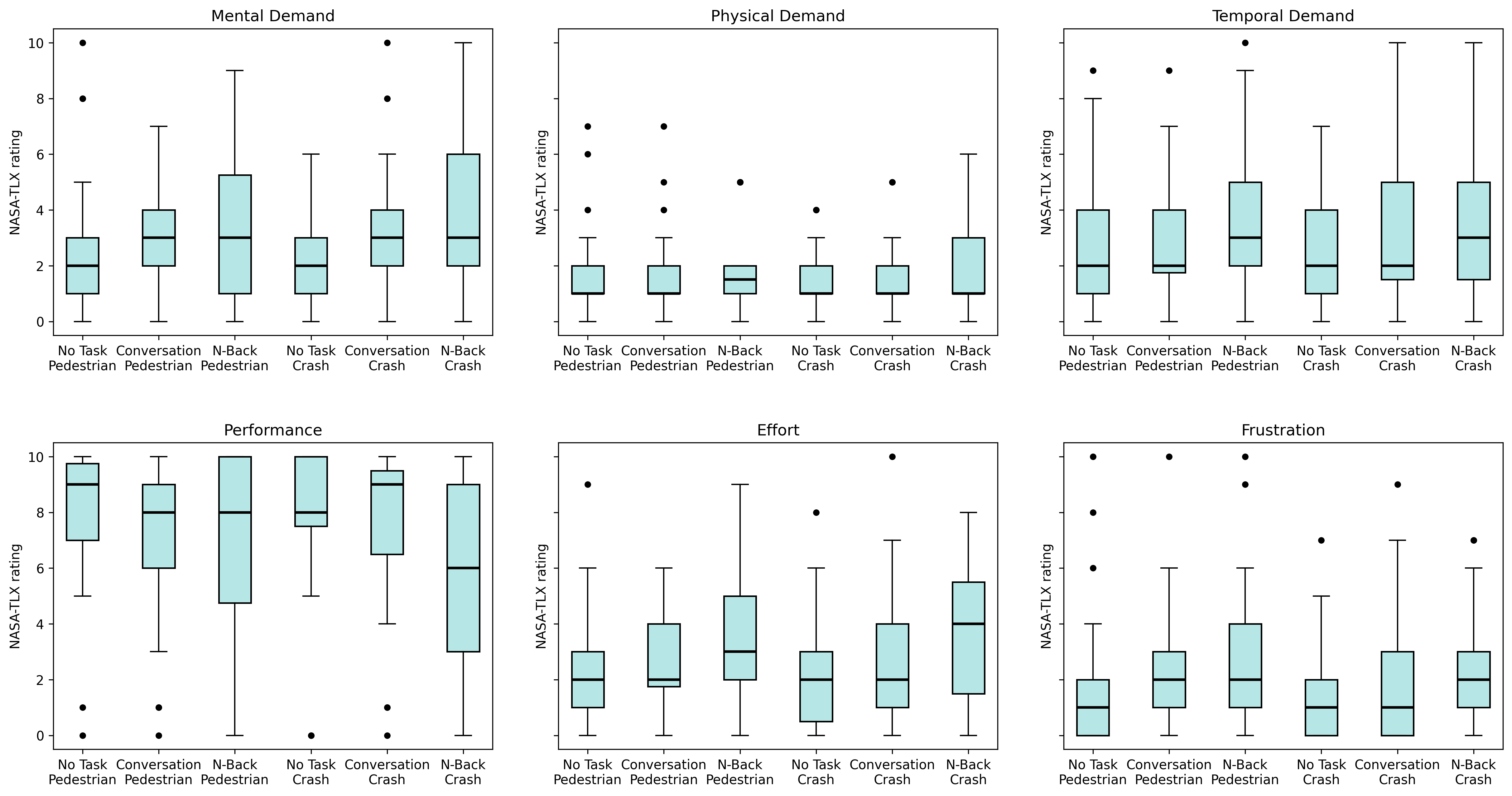}
\caption{Distribution of NASA-TLX
subscale ratings across conditions.}
\label{fig:tlx_boxplots}
\end{figure}

Figure~\ref{fig:tlx_boxplots} illustrates the distribution of NASA-TLX subscale ratings across conditions. Subjective workload was assessed using the NASA Task Load Index (NASA-TLX) and analyzed using linear mixed-effects models with Task, Event, and their interaction as fixed effects, and participant as a random intercept.

Table~\ref{tab:tlx_subscales} shows the corresponding mixed-effects model estimates for the NASA-TLX subscales. Across subscales, no significant main effects of Event or Task $\times$ Event interactions were observed, indicating that subjective workload did not differ reliably between Crash and Unexpected Pedestrian scenarios.
With respect to secondary task effects, the N-Back task was associated with significantly lower ratings of Mental Demand ($\beta = -1.26$, $t(149.45) = -2.55$, $p = 0.012$), Effort ($\beta = -1.32$, $t(150.00) = -3.29$, $p = 0.001$), and Frustration ($\beta = -1.06$, $t(149.03) = -2.75$, $p = 0.007$), as well as higher self-rated Performance ($\beta = 1.77$, $t(150.00) = 3.29$, $p = 0.001$), relative to the No-Task condition. The Conversation task was associated with a reduction in Effort ($\beta = -0.81$, $t(150.00) = -2.00$, $p = 0.047$), while its effects on other subscales were marginal or non-significant (e.g., Frustration, $p = 0.052$).
\small
\setlength{\tabcolsep}{4pt}
\setlength{\LTcapwidth}{\textwidth}
\setlength{\LTleft}{0pt}
\setlength{\LTright}{0pt}

\begin{longtable}{@{}p{3.7cm}ccccccc@{}}
\caption{Linear mixed-effects model results for NASA-TLX subscales.
Reference levels are No Task and Unexpected Pedestrian.}
\label{tab:tlx_subscales} \\
\hline
Term & Coef & StdErr & DF & $t$ & $p$ & 95\% CI$_L$ & 95\% CI$_U$ \\
\hline
\endfirsthead

\caption[]{Linear mixed-effects model results for NASA-TLX subscales (continued).} \\
\hline
Term & Coef & StdErr & DF & $t$ & $p$ & 95\% CI$_L$ & 95\% CI$_U$ \\
\hline
\endhead

\multicolumn{8}{l}{\textbf{Mental Demand}} \\
\hline
Intercept & 3.9355 & 0.4023 & 135.37 & 9.78 & $<$ .001 & 3.1471 & 4.7239 \\
Conversation & $-$0.7419 & 0.4909 & 149.12 & $-$1.51 & 0.133 & $-$1.7042 & 0.2203 \\
N-Back & $-$1.2643 & 0.4955 & 149.45 & $-$2.55 & 0.012 & $-$2.2356 & $-$0.2931 \\
Crash & 0.2903 & 0.4909 & 149.12 & 0.59 & 0.555 & $-$0.6719 & 1.2526 \\
Conversation $\times$ Crash & $-$0.7742 & 0.6943 & 149.12 & $-$1.12 & 0.267 & $-$2.1350 & 0.5866 \\
N-Back $\times$ Crash & $-$1.0905 & 0.6975 & 149.29 & $-$1.56 & 0.120 & $-$2.4577 & 0.2767 \\

\hline
\multicolumn{8}{l}{\textbf{Physical Demand}} \\
\hline
Intercept & 1.7097 & 0.2674 & 80.49 & 6.39 & $<$ .001 & 1.1855 & 2.2338 \\
Conversation & $-$0.0968 & 0.2682 & 150.00 & $-$0.36 & 0.719 & $-$0.6224 & 0.4288 \\
N-Back & $-$0.0323 & 0.2682 & 150.00 & $-$0.12 & 0.904 & $-$0.5579 & 0.4933 \\
Crash & 0.1290 & 0.2682 & 150.00 & 0.48 & 0.631 & $-$0.3966 & 0.6546 \\
Conversation $\times$ Crash & $-$0.1613 & 0.3793 & 150.00 & $-$0.43 & 0.671 & $-$0.9046 & 0.5820 \\
N-Back $\times$ Crash & $-$0.6129 & 0.3793 & 150.00 & $-$1.62 & 0.108 & $-$1.3562 & 0.1304 \\

\hline
\multicolumn{8}{l}{\textbf{Temporal Demand}} \\
\hline
Intercept & 3.6129 & 0.4133 & 137.76 & 8.74 & $<$ .001 & 2.8028 & 4.4230 \\
Conversation & $-$0.8387 & 0.5070 & 150.00 & $-$1.65 & 0.100 & $-$1.8324 & 0.1550 \\
N-Back & $-$0.8710 & 0.5070 & 150.00 & $-$1.72 & 0.088 & $-$1.8646 & 0.1227 \\
Crash & 0.0000 & 0.5070 & 150.00 & 0.00 & 1.000 & $-$0.9937 & 0.9937 \\
Conversation $\times$ Crash & $-$0.1290 & 0.7170 & 150.00 & $-$0.18 & 0.857 & $-$1.5343 & 1.2762 \\
N-Back $\times$ Crash & $-$0.7419 & 0.7170 & 150.00 & $-$1.03 & 0.302 & $-$2.1472 & 0.6633 \\

\hline
\multicolumn{8}{l}{\textbf{Performance}} \\
\hline
Intercept & 6.1290 & 0.4906 & 100.87 & 12.49 & $<$ .001 & 5.1674 & 7.0907 \\
Conversation & 1.0323 & 0.5392 & 150.00 & 1.91 & 0.057 & $-$0.0246 & 2.0891 \\
N-Back & 1.7742 & 0.5392 & 150.00 & 3.29 & 0.001 & 0.7173 & 2.8311 \\
Crash & $-$0.5484 & 0.5392 & 150.00 & $-$1.02 & 0.311 & $-$1.6053 & 0.5085 \\
Conversation $\times$ Crash & 1.2581 & 0.7626 & 150.00 & 1.65 & 0.101 & $-$0.2366 & 2.7527 \\
N-Back $\times$ Crash & 0.5806 & 0.7626 & 150.00 & 0.76 & 0.448 & $-$0.9140 & 2.0753 \\

\hline
\multicolumn{8}{l}{\textbf{Effort}} \\
\hline
Intercept & 3.5484 & 0.3550 & 109.66 & 9.99 & $<$ .001 & 2.8525 & 4.2442 \\
Conversation & $-$0.8065 & 0.4022 & 150.00 & $-$2.00 & 0.047 & $-$1.5948 & $-$0.0181 \\
N-Back & $-$1.3226 & 0.4022 & 150.00 & $-$3.29 & 0.001 & $-$2.1110 & $-$0.5342 \\
Crash & 0.0645 & 0.4022 & 150.00 & 0.16 & 0.873 & $-$0.7239 & 0.8529 \\
Conversation $\times$ Crash & $-$0.6452 & 0.5689 & 150.00 & $-$1.13 & 0.259 & $-$1.7601 & 0.4698 \\
N-Back $\times$ Crash & $-$0.6129 & 0.5689 & 150.00 & $-$1.08 & 0.283 & $-$1.7279 & 0.5021 \\

\hline
\multicolumn{8}{l}{\textbf{Frustration}} \\
\hline
Intercept & 2.9677 & 0.3956 & 76.00 & 7.50 & $<$ .001 & 2.1923 & 3.7432 \\
Conversation & $-$0.7635 & 0.3903 & 149.18 & $-$1.96 & 0.052 & $-$1.5285 & 0.0015 \\
N-Back & $-$1.0645 & 0.3866 & 149.03 & $-$2.75 & 0.007 & $-$1.8222 & $-$0.3069 \\
Crash & $-$0.4839 & 0.3866 & 149.03 & $-$1.25 & 0.213 & $-$1.2415 & 0.2738 \\
Conversation $\times$ Crash & $-$0.0107 & 0.5493 & 149.11 & $-$0.02 & 0.984 & $-$1.0874 & 1.0660 \\
N-Back $\times$ Crash & $-$0.3226 & 0.5467 & 149.03 & $-$0.59 & 0.556 & $-$1.3941 & 0.7489 \\

\hline
\end{longtable}

\subsection{Electrodermal Activity (EDA)}

Linear mixed-effects models were fitted for tonic activity, indexed by skin conductance level (SCL), phasic activity, indexed by skin conductance response (SCR) amplitude, SCR count, SCR rate, and total activation, indexed by the area under the curve (AUC), with Task, Event type, and their interaction as fixed effects and Subject as a random intercept.

Tonic activity (measured as SCL), phasic activity (measured as SCR amplitude), and SCR rate did not show significant effects of Task, Event type, or their interaction (see Appendix~\ref{app:eda_full} for full model results and corresponding figures), indicating that baseline arousal and the magnitude of individual phasic responses remained stable across conditions.

Significant effects of Task were observed for SCR count and AUC. As shown in Table~\ref{tab:lme_scr_count}, SCR count was lower during both the N-Back ($\beta = -2.91$, $t(157.76) = -3.28$, $p = 0.001$) and Conversation ($\beta = -1.90$, $t(157.67) = -2.17$, $p = 0.032$) conditions relative to the No Task baseline, indicating fewer discrete autonomic responses under cognitive load.

\begin{figure}[htbp]
\centering

\begin{subfigure}[b]{0.48\linewidth}
    \centering
    \includegraphics[width=\linewidth]{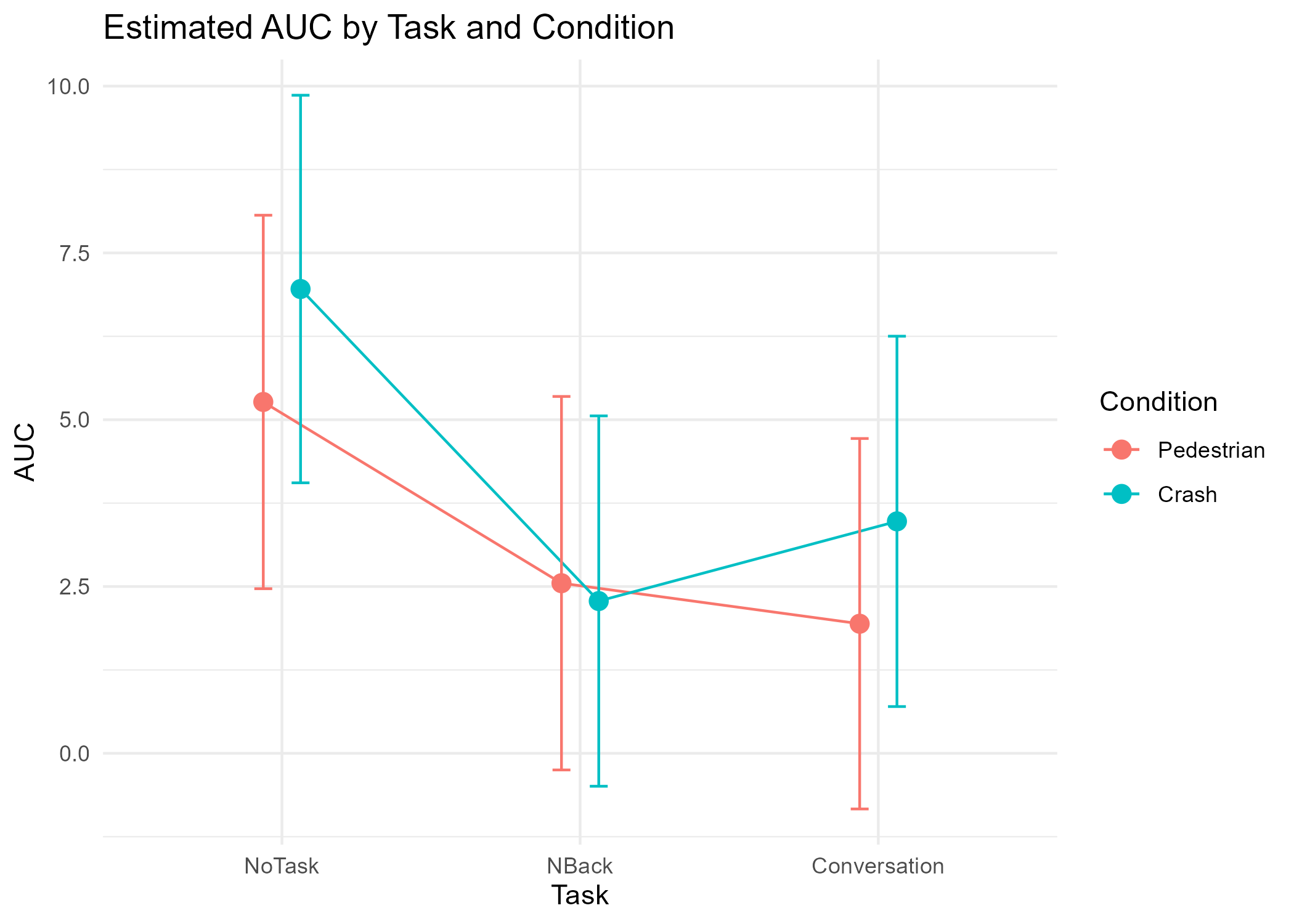}
    \caption{AUC}
    \label{fig:auc}
\end{subfigure}
\hfill
\begin{subfigure}[b]{0.48\linewidth}
    \centering
    \includegraphics[width=\linewidth]{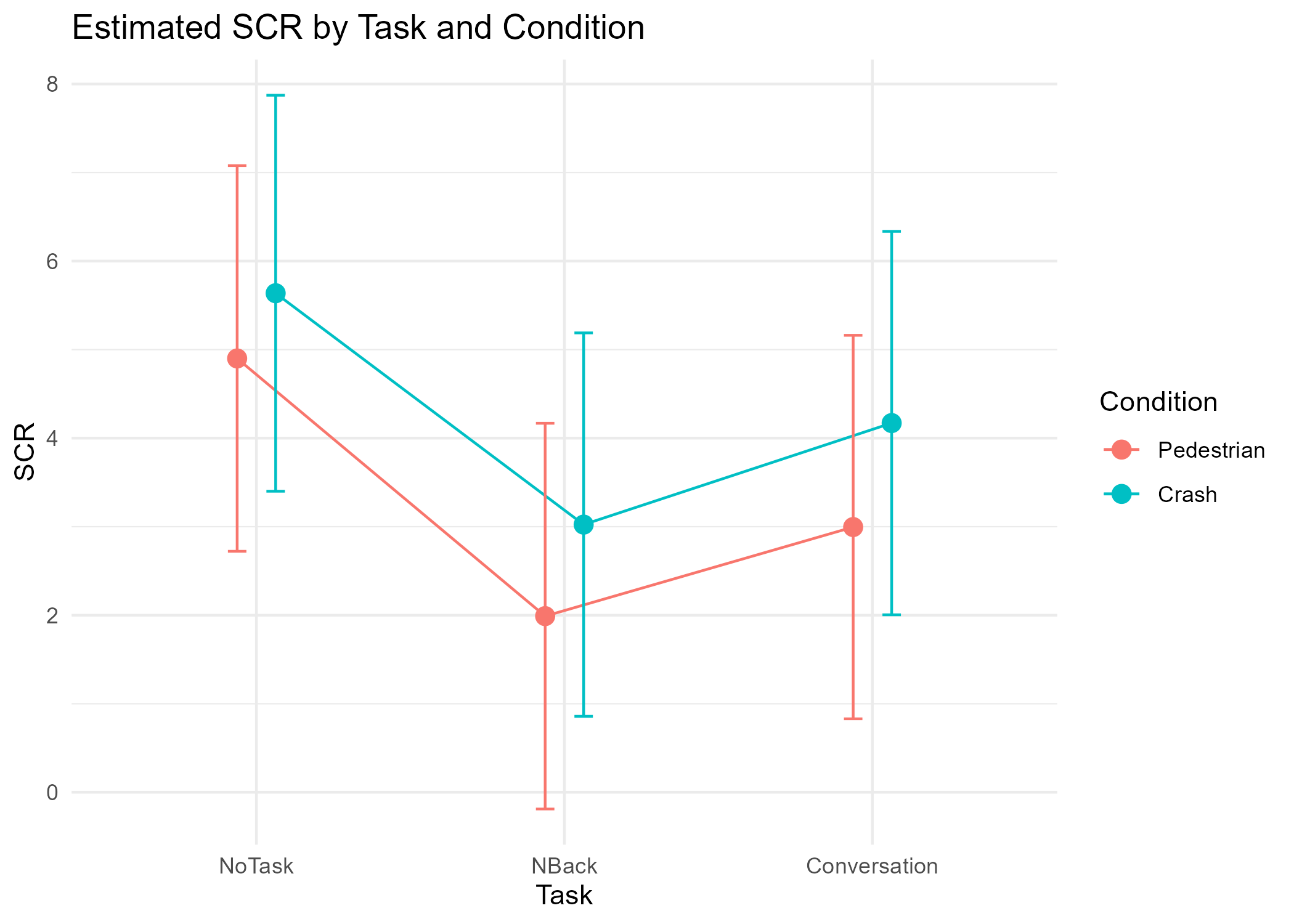}
    \caption{SCR count}
    \label{fig:scr}
\end{subfigure}

\caption{Estimated marginal means of (a) AUC and (b) SCR count across task and event conditions. Error bars represent 95\% confidence intervals.}

\label{fig:eda_sig}

\end{figure}

\begin{table}[H]
\centering
\caption{Linear mixed-effects model results for SCR count.
Reference levels are No Task and Unexpected Pedestrian.}
\label{tab:lme_scr_count}

\small
\setlength{\tabcolsep}{4pt}

\begin{adjustbox}{max width=\linewidth}
\begin{tabular}{lccccccc}
\hline
Term & Coef & StdErr & DF & $t$ & $p$ & 95\% CI$_{lower}$ & 95\% CI$_{upper}$ \\
\hline
Intercept & 4.90 & 1.07 & 64.38 & 4.56 & $<$ .001 & 2.79 & 7.01 \\
N-Back & $-$2.91 & 0.89 & 157.76 & $-$3.28 & 0.001 & $-$4.65 & $-$1.17 \\
Conversation & $-$1.90 & 0.88 & 157.67 & $-$2.17 & 0.032 & $-$3.63 & $-$0.18 \\
Crash & 0.74 & 0.92 & 157.96 & 0.80 & 0.425 & $-$1.07 & 2.54 \\
N-Back $\times$ Crash & 0.30 & 1.27 & 157.92 & 0.23 & 0.816 & $-$2.20 & 2.79 \\
Conversation $\times$ Crash & 0.44 & 1.27 & 157.89 & 0.35 & 0.730 & $-$2.05 & 2.92 \\
\hline
\end{tabular}
\end{adjustbox}
\end{table}

A similar pattern was observed for AUC, as shown in Table~\ref{tab:lme_auc}, with reduced values in both the N-Back ($\beta = -2.72$, $t(158.16) = -2.01$, $p = 0.046$) and Conversation ($\beta = -3.32$, $t(158.01) = -2.48$, $p = 0.014$) conditions relative to the No Task baseline, indicating reduced overall electrodermal activation under cognitive load.

No main effect of Event type or Task $\times$ Event interaction were observed across EDA metrics, indicating that electrodermal responses were not strongly modulated by hazard context.

\begin{table}[H]
\centering
\caption{Linear mixed-effects model results for EDA AUC. 
Reference levels are No Task and Unexpected Pedestrian.}
\label{tab:lme_auc}

\small
\setlength{\tabcolsep}{4pt}

\begin{adjustbox}{max width=\linewidth}
\begin{tabular}{lccccccc}
\hline
Term & Coef & StdErr & DF & $t$ & $p$ & 95\% CI$_{lower}$ & 95\% CI$_{upper}$ \\
\hline
Intercept & 5.27 & 1.39 & 85.89 & 3.80 & $<$ .001 & 2.55 & 7.98 \\
N-Back & $-$2.72 & 1.35 & 158.16 & $-$2.01 & 0.046 & $-$5.37 & $-$0.06 \\
Conversation & $-$3.32 & 1.34 & 158.01 & $-$2.48 & 0.014 & $-$5.95 & $-$0.69 \\
Crash & 1.69 & 1.41 & 158.48 & 1.20 & 0.230 & $-$1.06 & 4.45 \\
N-Back $\times$ Crash & $-$1.96 & 1.95 & 158.42 & $-$1.01 & 0.315 & $-$5.77 & 1.85 \\
Conversation $\times$ Crash & $-$0.16 & 1.94 & 158.38 & $-$0.08 & 0.935 & $-$3.96 & 3.64 \\
\hline
\end{tabular}
\end{adjustbox}
\end{table}

Figure~\ref{fig:eda_sig} shows that both SCR count and AUC decrease from the No Task condition to the secondary task conditions, with the largest reduction observed in the N-Back condition. While Crash trials tend to show higher overall values than Unexpected Pedestrian trials, the magnitude of the task-related reduction is comparable across event types.

\subsection{Heart Rate Variability (HRV)}

To examine autonomic responses during takeover events, HRV was analyzed using the root mean square of successive differences. RMSSD values were analyzed using a linear mixed-effects model with Task (No Task, Conversation, N-Back), Event type (Unexpected Pedestrian, Crash), and their interaction as fixed effects, and a random intercept for Subject to account for repeated measures.

Table~\ref{tab:rmssd_lmm} summarizes the fixed-effect estimates from the model. Overall, RMSSD did not differ significantly as a function of secondary task engagement or event type. Relative to the No Task condition during Unexpected Pedestrian events, neither the N-Back task ($\beta = -7.94$, $t(157.93) = -0.43$, $p = 0.666$) nor the Conversation task ($\beta = 2.15$, $t(157.75) = 0.12$, $p = 0.906$) was associated with a significant change in RMSSD. Similarly, no main effect of Event type was observed, with Crash events not differing significantly from Unexpected Pedestrian events ($\beta = 15.08$, $t(158.30) = 0.79$, $p = 0.430$).

No significant Task $\times$ Event interaction effects were detected. Specifically, the interaction between N-Back and Crash events ($\beta = -30.18$, $t(158.22) = -1.14$, $p = 0.255$) and the interaction between Conversation and Crash events ($\beta = 21.27$, $t(158.18) = 0.81$, $p = 0.420$) did not reach statistical significance, indicating that the relationship between secondary task engagement and RMSSD did not differ reliably between hazard contexts.

\begin{figure}[htbp]
\centering
\includegraphics[width=0.9\linewidth]{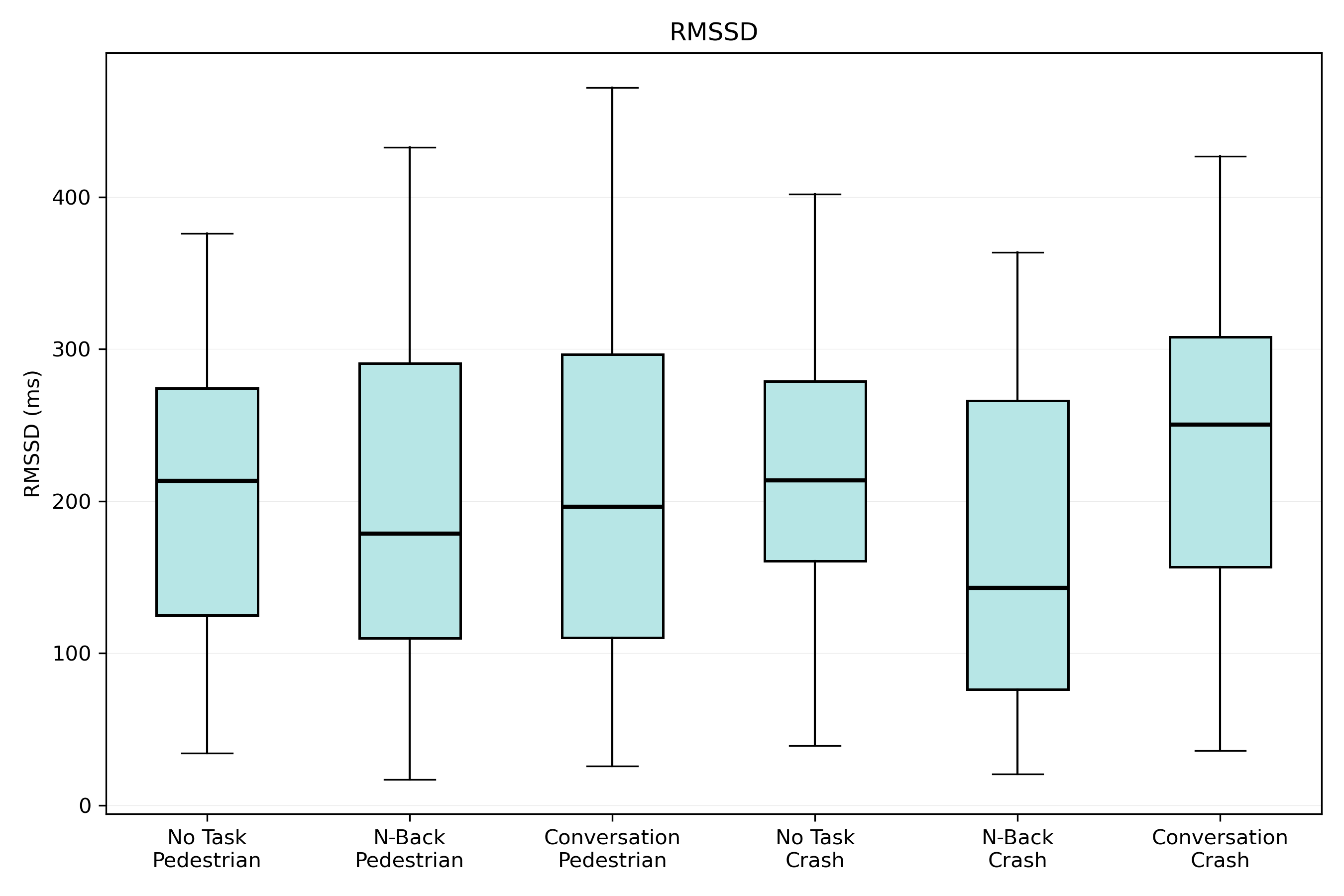}
\caption{Distribution of RMSSD during the takeover window across task and hazard conditions.}
\label{fig:rmssd}
\end{figure}

Figure~\ref{fig:rmssd} illustrates the model-estimated marginal means of RMSSD across task and hazard conditions. Although numerical differences are visible across conditions, these variations were not statistically significant, suggesting that autonomic cardiac regulation, as indexed by RMSSD, remained relatively stable across levels of secondary task engagement and hazard type during takeover execution.

\begin{table}[htbp]
\centering
\caption{Linear mixed-effects model results for RMSSD. Reference levels are No Task and Unexpected Pedestrian.}
\label{tab:rmssd_lmm}

\small
\setlength{\tabcolsep}{4pt}

\begin{adjustbox}{max width=\linewidth}
\begin{tabular}{lccccccc}
\hline
Term & Coef & StdErr & DF & $t$ & $p$ & 95\% CI$_{lower}$ & 95\% CI$_{upper}$ \\
\hline
Intercept & 194.99 & 18.09 & 92.77 & 10.78 & 4.88e$-$18 & 159.53 & 230.45 \\
Conversation & 2.15 & 18.21 & 157.75 & 0.12 & 0.906 & $-$33.53 & 37.84 \\
N-Back & $-$7.94 & 18.37 & 157.93 & $-$0.43 & 0.666 & $-$43.95 & 28.07 \\
Crash & 15.08 & 19.08 & 158.30 & 0.79 & 0.430 & $-$22.31 & 52.46 \\
Conversation $\times$ Crash & 21.27 & 26.30 & 158.18 & 0.81 & 0.420 & $-$30.27 & 72.81 \\
N-Back $\times$ Crash & $-$30.18 & 26.41 & 158.22 & $-$1.14 & 0.255 & $-$81.93 & 21.58 \\
\hline
\end{tabular}
\end{adjustbox}
\end{table}

To further characterize condition-specific patterns, follow-up analyses were conducted separately within each hazard context. Although the Task $\times$ Event interaction was not statistically significant in the full mixed-effects model, stratified analyses were performed to examine within-condition differences in RMSSD across task conditions. These analyses are descriptive in nature and are intended to aid interpretation of the overall pattern of results.

Within Crash events, a significant main effect of Task on RMSSD was observed (Type III ANOVA: $F(2, 60.73) = 6.19$, $p = 0.0036$; Table~\ref{tab:rmssd_crash_anova}). Estimated marginal means indicated lower RMSSD during the N-Back task compared to Conversation (Tukey-adjusted $p = 0.0028$; Table~\ref{tab:rmssd_crash_emm}). Differences between No Task and N-Back ($p = 0.0826$) and between No Task and Conversation ($p = 0.5246$) did not reach statistical significance.

\begin{table}[htbp]
\centering
\caption{Type III analysis of variance for RMSSD during Crash events.}
\label{tab:rmssd_crash_anova}

\small
\setlength{\tabcolsep}{5pt}

\begin{adjustbox}{max width=\linewidth}
\begin{tabular}{lcccccc}
\hline
Effect & Sum Sq & Mean Sq & Num DF & Den DF & $F$ & $p$ \\
\hline
Task & 67022 & 33511 & 2 & 60.73 & 6.19 & 0.0036 \\
\hline
\end{tabular}
\end{adjustbox}
\end{table}

\begin{table}[H]
\centering
\caption{Estimated marginal means and Tukey-adjusted pairwise comparisons of RMSSD across task conditions during Crash events.}
\label{tab:rmssd_crash_emm}

\small
\setlength{\tabcolsep}{5pt}

\begin{adjustbox}{max width=\linewidth}
\begin{tabular}{lccccc}
\hline
Task & Emmean & SE & DF & 95\% CI$_{lower}$ & 95\% CI$_{upper}$ \\
\hline
No Task & 212 & 18.5 & 74.2 & 175 & 248 \\
N-Back & 169 & 17.4 & 67.0 & 135 & 204 \\
Conversation & 233 & 17.4 & 67.0 & 198 & 267 \\
\hline
\end{tabular}
\end{adjustbox}

\vspace{0.8em}

\small
\setlength{\tabcolsep}{5pt}

\begin{adjustbox}{max width=\linewidth}
\begin{tabular}{lccccc}
\hline
Contrast & Estimate & SE & DF & $t$ & $p$ \\
\hline
No Task -- N-Back & 42 & 19.3 & 60.3 & 2.18 & 0.0826 \\
No Task -- Conversation & $-$21 & 19.3 & 60.3 & $-$1.09 & 0.5246 \\
N-Back -- Conversation & $-$63 & 18.2 & 58.7 & $-$3.46 & 0.0028 \\
\hline
\end{tabular}
\end{adjustbox}
\end{table}

Within Unexpected Pedestrian events, RMSSD did not differ significantly across task conditions. Estimated marginal means and Tukey-adjusted pairwise comparisons revealed no significant differences between No Task, N-Back, and Conversation conditions (Table~\ref{tab:rmssd_pedestrian_emm}).

\begin{table}[H]
\centering
\caption{Estimated marginal means and Tukey-adjusted pairwise comparisons of RMSSD across task conditions during Unexpected Pedestrian events.}
\label{tab:rmssd_pedestrian_emm}

\small
\setlength{\tabcolsep}{5pt}

\begin{adjustbox}{max width=\linewidth}
\begin{tabular}{lccccc}
\hline
Task & Emmean & SE & DF & 95\% CI$_{lower}$ & 95\% CI$_{upper}$ \\
\hline
No Task & 196 & 19.2 & 69.3 & 157 & 234 \\
N-Back & 187 & 19.2 & 69.3 & 149 & 226 \\
Conversation & 198 & 19.0 & 67.9 & 160 & 236 \\
\hline
\end{tabular}
\end{adjustbox}

\vspace{0.8em}

\small
\setlength{\tabcolsep}{5pt}

\begin{adjustbox}{max width=\linewidth}
\begin{tabular}{lccccc}
\hline
Contrast & Estimate & SE & DF & $t$ & $p$ \\
\hline
No Task -- N-Back & 8.39 & 20.2 & 62.3 & 0.42 & 0.9092 \\
No Task -- Conversation & $-$2.06 & 20.0 & 62.0 & $-$0.10 & 0.9942 \\
N-Back -- Conversation & $-$10.45 & 20.0 & 62.0 & $-$0.52 & 0.8603 \\
\hline
\end{tabular}
\end{adjustbox}
\end{table}

\subsection{Functional near-infrared spectroscopy (fNIRS)}

To evaluate task-related differences in cortical activation, we conducted mixed-effects analyses. Specifically, a linear mixed-effects model (LME) was fitted to the entire dataset to assess the overall effect of condition on HbO concentration changes across both participants, conditions (tasks), alongside event (first, 'Unexpected Pedestrian', and second, 'Crash'), and channels. In this model, the Condition-event was included as a fixed effect, and Subject and Channel were modeled as random intercepts to account for inter-individual and inter-channel variability, respectively. Post-hoc pairwise contrasts between conditions were computed based on the estimated fixed effects of the model.

The overall model revealed a significant main effect of Condition on HbO concentration changes (F(5, 1044) = 3.56, $p = 0.005$), indicating that cortical activation differed across experimental conditions. Table~\ref{fNIRS_LME} shows the corresponding mixed-effects model estimates. The fixed effects analysis revealed that, compared to the reference condition (No Task $\times$ Unexpected Pedestrian), several conditions such as Conversation $\times$ Unexpected Pedestrian and N-Back $\times$ Crash showed significant reductions in HbO concentration (estimates = -0.25, $p = 0.0039$ and -0.21, $p = 0.0169$, respectively). In contrast, No Task $\times$ Crash and Conversation $\times$ Crash did not show significant differences ($p > 0.05$ for both). Model fit indices (AIC = 2641.42; BIC = 2686.03) suggested an adequate fit, with random intercept variance mainly attributable to subjects (SD = 0.30) and minimal variance at the channel level (SD $\approx$ 0.048), reflecting consistent spatial patterns across brain regions during the tasks.

The observed reduction or muted increase in HbO levels during the N-Back and Conversation tasks while driving is likely a result of competing cognitive demands. Although cognitive tasks such as the N-Back are typically associated with increased HbO in controlled settings due to elevated neural activity, the presence of the driving task introduces additional cognitive load. This shared demand on the brain, particularly in areas such as the prefrontal cortex involved in both driving and task performance, may limit the brain's ability to increase HbO. As a result, no significant increase, or even a reduction in HbO, may be observed compared to non-driving baselines.

In summary, while tasks like the N-Back and Conversation are generally associated with increased HbO in non-driving contexts, the dual-tasking environment during driving likely leads to a reduction or attenuated increase in HbO. The competing demands from the driving task may restrict the brain's capacity to allocate sufficient oxygen to support the additional cognitive load imposed by the N-Back or Conversation tasks, thereby resulting in a blunted or reduced HbO response.



\begin{table}[H]
\centering
\caption{Linear mixed-effects model results for HbO \(\sim\) Condition + (1|Subject) + (1|Channel).
Reference levels are No Task First (Unexpected Pedestrian).}
\label{fNIRS_LME}

\small
\setlength{\tabcolsep}{4pt}

\begin{adjustbox}{max width=\linewidth}
\begin{tabular}{lccccccc}
\hline
Term & Coef & StdErr & DF & $t$ & $p$ & 95\% CI$_{lower}$ & 95\% CI$_{upper}$ \\
\hline
Intercept & 0.1816 & 0.0848 & 1044 & 2.1416 & 0.0325 & 0.0152 & 0.3481 \\
Conversation & $-$0.2520 & 0.0872 & 1044 & $-$2.8889 & 0.0039 & $-$0.4232 & $-$0.0808 \\
N-Back & $-$0.0609 & 0.0871 & 1044 & $-$0.6986 & 0.4849 & $-$0.2318 & 0.1101 \\
Crash & 0.0230 & 0.0893 & 1044 & 0.2577 & 0.7967 & $-$0.1522 & 0.1982 \\
Conversation $\times$ Crash & $-$0.0306 & 0.0874 & 1044 & $-$0.3506 & 0.7259 & $-$0.2021 & 0.1408 \\
N-Back $\times$ Crash & $-$0.2095 & 0.0875 & 1044 & $-$2.3935 & 0.0169 & $-$0.3812 & $-$0.0377 \\
\hline
\end{tabular}
\end{adjustbox}

\end{table}

\section{Discussion} \label{sec:discussion}
The findings of this study indicate that takeover performance in semi-automated driving is primarily governed by hazard context rather than secondary cognitive load. Across behavioral measures, including execution duration, speed regulation, braking patterns, and final maneuver strategies, hazard type consistently explained substantial variation in driver responses, whereas differences across secondary task conditions were comparatively limited. In contrast, secondary task engagement showed more variable effects on drivers’ internal states, with evidence of task-related differences in subjective workload, physiological responses, and neural measures. Together, these findings suggest that observable takeover behavior and underlying cognitive state do not necessarily align during automated driving transitions. In the following sections, we first discuss the counterintuitive patterns observed across multimodal measures of driver state under secondary-task engagement, and then examine how hazard context shaped observable takeover behavior and control strategies during the takeover process.

Across subjective, physiological, and neural measures, secondary task engagement produced a pattern that was counterintuitive relative to the expected neuro and psycho-physiological effects of cognitive load, but largely consistent across modalities. Subjective workload ratings were generally low across conditions and showed limited sensitivity to secondary task engagement, with no significant effect of hazard type. In particular, the N-Back condition was associated with lower perceived mental demand, effort, and frustration, along with higher self-rated performance, than the no-task condition. This finding appears counterintuitive given that working-memory tasks are typically expected to increase perceived cognitive demand, and prior work suggests that subjective workload measures such as NASA-TLX can capture variations in cognitive demand that are not always reflected in observable driving performance \citep{liu2024safety, muller2021effects}. A similar pattern was observed in neural activity. HbO was significantly reduced during the Conversation condition in the Unexpected Pedestrian scenario and during the N-Back condition in the Crash scenario, relative to the No Task condition in the Unexpected Pedestrian scenario. This finding contrasts with the typical expectation that cognitive tasks such as N-Back increase prefrontal activation in non-driving contexts, reflecting greater cognitive demand and executive engagement \citep{lohani2019review, unni2017assessing, hergeth2016keep}. Electrodermal activity showed the same unexpected direction. SCR count and AUC were reduced under secondary task conditions, indicating lower physiological responsiveness. Because electrodermal activity is commonly interpreted as a marker of sympathetic arousal that increases under cognitive load and distraction \citep{halin2025electrodermal, boffet2025detection}, secondary tasks would normally be expected to produce more frequent and larger electrodermal responses relative to the no-task baseline. However, our results showed reduced rather than increased electrodermal responses. 


This convergence across subjective workload, electrodermal activity, and fNIRS measures suggests that the no-task condition with pedestrian may have involved a distinct form of attentional demand rather than a purely low-workload baseline. During automated driving, drivers in the no-task condition monitored the automation without active vehicle control or an explicit secondary task. Prior work on partially automated driving shows that this type of passive supervisory role, while appearing low in demand, is associated with cognitive underload, vigilance decrement, and mind wandering, all of which reflect a diffuse and unstructured attentional state rather than genuine cognitive rest \citep{mcwilliams2021underload, mishler2024boring}. Crucially, this state is not neutral: drivers in passive monitoring conditions must remain alert to rare and unpredictable safety-critical events, which may generate a sustained form of anticipatory arousal that is distinct from the arousal produced by an active cognitive task. By contrast, the N-Back and conversation tasks provided a structured attentional focus during the automated driving period, which may reflect a more directed form of cognitive engagement. This could explain why secondary task conditions were associated with lower rather than higher subjective workload, electrodermal activity, and prefrontal activation relative to the no-task baseline. This interpretation is consistent with the underload account, which predicts that secondary tasks can mitigate the vigilance decrement during passive monitoring by increasing task engagement rather than depleting attentional resources \citep{mishler2024boring}. 

An additional explanation relates to strategic task prioritization. Drivers anticipating a safety-critical takeover may selectively reduce engagement with an assigned secondary task and redirect attention toward the road. Research on driver self-regulation has shown that drivers tend to engage in secondary tasks primarily during low-demand situations and may continue monitoring the driving scene rather than fully committing to an imposed task \citep{wandtner2018secondary, gremillion2016analysis, lee2014dynamics}. If participants in the N-Back and conversation conditions similarly deprioritized the secondary task, these conditions may have functioned more like active road-monitoring states, helping explain their lower physiological and neural responses relative to the no-task baseline.

A different pattern emerged when considering observable takeover behavior. Unlike the internal-state measures, which showed counterintuitive and task-dependent effects, behavioral responses were consistently shaped by the characteristics of the hazard context itself. A key reason for this pattern lies in the nature of the hazard context itself. The differences observed between pedestrian and crash scenarios reflect how drivers interpret and respond to uncertainty during the takeover process. Pedestrian hazards introduce a dynamic and less predictable situation, requiring continuous monitoring of the pedestrian’s movement while drivers regain control and determine how to respond. Under these conditions, it becomes more difficult to commit to a single, well-defined action, which leads to longer and more variable control adjustments. In contrast, the crash hazard represents a stationary and spatially predictable obstacle, allowing drivers to more quickly recognize the situation and execute an avoidance plan, resulting in shorter and more stable responses. Similar distinctions between dynamic and static hazards have been reported in prior work, where increased uncertainty is associated with more variable control behavior, while predictable obstacles allow for more structured responses \citep{duan2025analysis, gold2016taking}. From a cognitive perspective, this pattern aligns with situation awareness, in which drivers must perceive, comprehend, and anticipate the state of the environment to guide their actions \citep{mcdonald2019toward}. Overall, these results suggest that how drivers interpret the hazard plays an important role in shaping takeover behavior.

Secondary task engagement showed a limited influence on observable takeover behavior in this study. Rather than producing clear differences in overall driving performance, its effects were subtle and became noticeable mainly in how drivers adjusted control from moment to moment. In particular, subtle differences were observed in measures such as speed and deceleration variability, most clearly during the conversation task and especially in the crash scenario, where driver responses were otherwise more stable. In contrast, these effects were not clearly visible in the pedestrian scenario, where the dynamic and uncertain nature of the hazard already required continuous adjustment and introduced higher baseline variability. In such conditions, additional effects of the secondary task are likely masked rather than absent. This pattern aligns with prior work indicating that, although non-driving-related tasks can affect takeover performance, their observable effects are not always consistently reflected across behavioral measures and depend on task characteristics and context, represented by hazard type in the current study \cite{weaver2022systematic}. One possible explanation is that the takeover request prompts drivers to rapidly re-engage with the driving task and prioritize vehicle control, allowing them to stabilize their behavior during the maneuver.

Our findings highlight the need to consider both environmental context and drivers’ psychophysiological state when evaluating takeover performance. Hazard characteristics appear to shape observable control behavior, whereas secondary task engagement may influence the internal processes through which drivers maintain readiness during automation. This distinction suggests that stable driving performance does not necessarily indicate low cognitive demand, and that behavioral measures alone may provide an incomplete view of takeover readiness. A multimodal perspective that integrates behavioral, subjective, neural, and physiological indicators can therefore offer a more comprehensive understanding of takeover dynamics. The importance of multimodal sensing for characterizing driver readiness, while motivated in our study by the observed dissociation between behavioral performance and internal state, is consistent with arguments made in prior work. Researchers have noted that although drivers sometimes show no observable variation at the performance level, their cognitive and emotional states may be substantially altered, and that psychophysiological measures provide sensitivity and specificity to capture the internal states that behavioral indicators alone cannot reveal \citep{du2020psychophysiological, deng2024analysis}. More broadly, the case for integrating physiological data into driver monitoring systems has been advanced precisely because assessments of takeover readiness based on behavior alone risk missing the underlying state from which that behavior emerges \citep{coyne2023assessing, pakdamanian2021deeptake}.

\subsection{Implications for Driver Monitoring and Automated Driving System Design}

From a design perspective, incorporating context-aware strategies that account for both hazard characteristics and driver condition may better support safe and adaptive transitions between automated and manual control. The finding that hazard type is the key factor that impacts observable takeover behavior suggests that automated systems should be capable of classifying the nature of an upcoming hazard and tailoring the takeover request accordingly — for instance, issuing earlier or more urgent alerts for dynamic pedestrian hazards, which were shown to produce longer and more variable maneuvers, compared to static crash scenarios where drivers responded more quickly and consistently. At the same time, the divergence between observable driving performance and internal cognitive state highlights a critical limitation of behavior-only monitoring approaches: a driver may appear to be performing adequately while simultaneously operating under elevated cognitive load, reduced physiological responsiveness, or altered neural activation. This underscores the need for driver monitoring systems that integrate multiple sensing modalities — including physiological signals such as EDA and HRV, neurophysiological measures such as fNIRS, and vehicle-based behavioral indicators to construct a more complete and accurate picture of driver readiness in real time.

The present findings also carry a less obvious but important implication for how driver readiness is conceptualized during the passive monitoring phase of semi-automated driving. Conventional assumptions in system design treat the no-task condition as a neutral or low-demand baseline from which drivers are best positioned to respond to a takeover request. The internal state findings of this study challenge that assumption. If passive monitoring involves its own form of attentional demand, as the convergent pattern across NASA-TLX, EDA, and fNIRS suggests, then the default state of a driver without an explicit task is not necessarily one of calm readiness. Instead, this state may be less structured and less directly anchored to the driving environment than a state supported by a defined, low-demand task. This has direct consequences for takeover system design. Keeping drivers in a pure supervisory role with no structured engagement may produce less predictable internal states and, by extension, less consistent takeover readiness than providing them with a controlled form of cognitive engagement during automated phases. Rather than treating secondary task prohibition as the safest default, future systems may benefit from considering structured, low-demand engagement as a means of maintaining more stable and recoverable driver states. This reframes the design problem: the goal is not simply to minimize distraction but to manage the quality of driver engagement during automation in ways that support reliable transitions to manual control.

Together, our findings point toward a future of human-automation collaboration in which vehicles are not merely passive recipients of driver input, but active partners that continuously sense, interpret, and respond to the human state of the driver to ensure safer and more reliable transitions between automated and manual control.

\section{Limitations and Future Work} \label{sec:limitation}

Several limitations should be considered when interpreting the findings of this study, particularly in relation to how the experimental design shapes the interpretation, scope, and broader implications of the observed behavioral and neurophysiological responses.

The study was conducted within a high-fidelity driving simulator environment. Although simulator-based experimentation enables precise control over hazardous events and allows identical takeover conditions to be presented safely across participants, simulator environments inevitably simplify many aspects of real-world driving. Real-world takeover situations involve richer sensory inputs, environmental uncertainty, emotional consequences, and motivational pressures that may alter driver attention, urgency perception, and response strategies. As a result, the takeover behaviors observed in the present study should not be interpreted as direct representations of naturalistic driving behavior, but rather as controlled behavioral responses under experimentally standardized conditions. This limitation is particularly important when considering the practical deployment of automated driving systems, since drivers in real-world environments may exhibit greater variability, delayed engagement, or more emotionally influenced responses than those observed in the simulator. At the same time, the controlled nature of the simulator was essential for isolating relationships between hazard context, secondary-task engagement, and multimodal driver responses that would be substantially more difficult to disentangle in uncontrolled field settings. Future studies should therefore examine whether the behavioral and neurophysiological patterns observed in controlled simulator environments remain stable under more naturalistic driving conditions. In particular, combining simulator-based experimentation with field operational tests, naturalistic driving studies, or mixed-reality experimental frameworks may help clarify how real-world environmental complexity, perceived risk, and driver emotional engagement influence takeover performance and cognitive state transitions during automated driving.


The hazard and secondary-task conditions examined in the experiment also represent only a subset of the broader spectrum of situations encountered during automated driving. Importantly, the observed differences between hazard conditions likely reflect the combined influence of multiple interacting scenario characteristics rather than the isolated effect of a single variable. The hazard scenarios differed simultaneously across dimensions such as motion characteristics, predictability, salience, urgency, and available response time. Consequently, the findings should be interpreted as reflecting integrated responses to distinct takeover contexts rather than definitive causal effects tied exclusively to pedestrian motion or crash presence. This distinction has important implications for how takeover behavior is conceptualized within automated driving research. Human responses during transitions of control are likely shaped by the interaction of multiple contextual features operating simultaneously, suggesting that future system design and evaluation frameworks may benefit from treating takeover scenarios as multidimensional behavioral environments rather than isolated hazard categories. Similarly, the selected secondary tasks primarily represented cognitive engagement during automation, whereas many real-world distractions involve visual–manual interaction, multitasking, social communication, or emotionally engaging content. The present findings therefore provide insight into how particular forms of cognitive engagement interact with takeover scenarios, but they should not be generalized to all forms of distraction during automated driving. One major direction of future work should be to investigate takeover behavior across a broader and more systematically varied set of hazard configurations and distraction types in order to better disentangle how specific contextual factors influence driver response. In particular, experimental designs that independently manipulate characteristics such as urgency, predictability, hazard salience, response time availability, and distraction modality may help clarify which combinations of factors most strongly shape cognitive engagement, takeover quality, and driver adaptation during automated driving. Expanding the range of secondary tasks to include visually demanding, emotionally engaging, and socially interactive activities may also provide a more ecologically valid understanding of how real-world distractions influence transitions of control in partially automated vehicles.


The sensing and analytical framework used in the study also constrains the interpretation of the observed cognitive and behavioral patterns. Behavioral responses were primarily inferred from simulator-derived vehicle dynamics, while neurophysiological interpretation relied on prefrontal cortical activity measured using a single fNIRS configuration. Although these measures captured important aspects of workload, attentional engagement, and driving response, they provide only a partial representation of the broader perceptual, cognitive, and motor processes underlying takeover performance. Takeover behavior likely emerges through distributed interactions involving situational awareness, visual perception, motor planning, and decision-making processes that extend beyond the regions and modalities captured in the present study. This limitation is important because it affects how confidently internal cognitive states can be inferred from the measured signals. For example, reductions or increases in prefrontal activation may reflect overlapping mechanisms related to cognitive engagement, attentional allocation, automation monitoring, or adaptive disengagement rather than a single interpretable construct. Future work integrating broader neural sensing, eye-tracking, multimodal physiological measures, and more granular behavioral indicators may therefore help develop more comprehensive models of driver state transitions during automated driving.

The generalizability of the findings is also influenced by participant characteristics and experimental structure. Human takeover behavior is known to vary substantially across factors such as age, driving experience, familiarity with automated systems, cognitive flexibility, and individual risk perception. In addition, the ordering of hazard scenarios was not fully randomized in the same manner as the secondary-task conditions, introducing the possibility of learning, anticipation, or adaptation effects across repeated exposures. This consideration is particularly important in the context of the present findings because the pedestrian hazard, which consistently appeared earlier, represented the more dynamic and behaviorally demanding scenario, producing longer and more variable takeover responses. Early exposure to this higher-uncertainty condition may have increased participant vigilance or preparedness during later crash events, potentially contributing to the faster and more stable responses observed in the crash condition. More broadly, repeated interaction with automation may gradually reshape driver expectations, vigilance, and trust calibration over time, meaning that the observed behavioral and neurophysiological patterns may reflect not only responses to the immediate scenarios themselves, but also adaptation to the experimental environment. Future studies employing larger and more heterogeneous samples, fully randomized exposure structures, and longitudinal designs may help clarify the stability and transferability of the patterns identified in the present work across populations and longer-term human–automation interactions.

\section{Conclusion} \label{sec:conclusion}

This study examined how hazard context and secondary task engagement relate to driver takeover behavior in a semi-automated driving environment. The results consistently show that hazard type is the primary factor shaping observable takeover behavior, with dynamic pedestrian hazards leading to longer, more variable control responses and crash scenarios producing shorter and more stable maneuvers. In contrast, secondary task engagement had a limited impact on objective driving performance, while subjective workload, physiological responses, and prefrontal cortical activation showed task-related differences that were not always reflected in vehicle-control behavior. This pattern indicates that behavioral stability during takeover does not necessarily reflect a uniform internal driver state, and that subjective, physiological, and neural responses may vary even when external performance appears unchanged. It also suggests that the no-task condition may not represent a neutral low-demand baseline, but rather a distinct passive monitoring state. These findings highlight the importance of jointly considering environmental context and driver state when evaluating takeover readiness and designing adaptive driver monitoring and takeover support systems in semi-automated vehicles.

\section{Acknowledgment}

The authors would like to thank the Villanova Institute for Research and Scholarship (VIRS) for supporting this research through the Research Catalyst Grant (RCG). The authors also thank the College of Engineering, Center for Human-Environmental Systems (CHES), and the Department of Civil and Environmental Engineering at Villanova University for supporting the driving simulator used in our study.

\appendix
\section{Supplementary Electrodermal Activity Results}
\label{app:eda_full}

Additional electrodermal activity (EDA) metrics are reported here to provide a complete account of the analyses. These include tonic activity (skin conductance level; SCL), phasic activity (skin conductance response amplitude; SCR amplitude), and SCR rate, which did not show significant effects in the main analysis. The corresponding linear mixed-effects model results and supplementary figures (Figure~\ref{fig:eda_supplementary}) are presented below.

\begin{table}[htbp]
\centering
\caption{Linear mixed-effects model results for tonic activity (SCL).
Reference levels are No Task and Unexpected Pedestrian.}
\label{tab:lme_tonic}

\small
\setlength{\tabcolsep}{4pt}

\begin{adjustbox}{max width=\linewidth}
\begin{tabular}{lccccccc}
\hline
Term & Coef & StdErr & DF & $t$ & $p$ & 95\% CI$_{lower}$ & 95\% CI$_{upper}$ \\
\hline
Intercept & 1.57 & 0.55 & 36.43 & 2.87 & 0.007 & 0.50 & 2.64 \\
N-Back & 0.11 & 0.16 & 157.06 & 0.73 & 0.464 & $-$0.19 & 0.42 \\
Conversation & 0.07 & 0.15 & 157.06 & 0.44 & 0.663 & $-$0.24 & 0.37 \\
Crash & $-$0.08 & 0.16 & 157.08 & $-$0.48 & 0.631 & $-$0.40 & 0.24 \\
N-Back $\times$ Crash & $-$0.10 & 0.22 & 157.08 & $-$0.45 & 0.652 & $-$0.54 & 0.34 \\
Conversation $\times$ Crash & $-$0.16 & 0.22 & 157.08 & $-$0.72 & 0.474 & $-$0.60 & 0.28 \\
\hline
\end{tabular}
\end{adjustbox}
\end{table}

\begin{table}[htbp]
\centering
\caption{Linear mixed-effects model results for phasic activity (SCR amplitude).
Reference levels are No Task and Unexpected Pedestrian.}
\label{tab:lme_phasic}

\small
\setlength{\tabcolsep}{4pt}

\begin{adjustbox}{max width=\linewidth}
\begin{tabular}{lccccccc}
\hline
Term & Coef & StdErr & DF & $t$ & $p$ & 95\% CI$_{lower}$ & 95\% CI$_{upper}$ \\
\hline
Intercept & 0.058 & 0.021 & 71.25 & 2.72 & 0.008 & 0.016 & 0.101 \\
N-Back & $-$0.005 & 0.019 & 157.89 & $-$0.28 & 0.777 & $-$0.043 & 0.032 \\
Conversation & $-$0.018 & 0.019 & 157.78 & $-$0.94 & 0.348 & $-$0.054 & 0.019 \\
Crash & 0.026 & 0.020 & 158.13 & 1.31 & 0.191 & $-$0.013 & 0.064 \\
N-Back $\times$ Crash & $-$0.032 & 0.027 & 158.08 & $-$1.17 & 0.244 & $-$0.085 & 0.022 \\
Conversation $\times$ Crash & 0.005 & 0.027 & 158.05 & 0.18 & 0.858 & $-$0.048 & 0.058 \\
\hline
\end{tabular}
\end{adjustbox}
\end{table}

\begin{table}[htbp]
\centering
\caption{Linear mixed-effects model results for SCR rate.
Reference levels are No Task and Unexpected Pedestrian.}
\label{tab:lme_scr_rate}

\small
\setlength{\tabcolsep}{4pt}

\begin{adjustbox}{max width=\linewidth}
\begin{tabular}{lccccccc}
\hline
Term & Coef & StdErr & DF & $t$ & $p$ & 95\% CI$_{lower}$ & 95\% CI$_{upper}$ \\
\hline
Intercept & 3.28 & 1.09 & 66.31 & 3.00 & 0.004 & 1.14 & 5.42 \\
N-Back & $-$0.54 & 0.92 & 157.79 & $-$0.58 & 0.562 & $-$2.34 & 1.27 \\
Conversation & 0.27 & 0.91 & 157.70 & 0.30 & 0.765 & $-$1.51 & 2.06 \\
Crash & 0.73 & 0.96 & 158.00 & 0.76 & 0.449 & $-$1.15 & 2.60 \\
N-Back $\times$ Crash & 0.50 & 1.32 & 157.96 & 0.38 & 0.707 & $-$2.09 & 3.09 \\
Conversation $\times$ Crash & 0.76 & 1.32 & 157.93 & 0.58 & 0.563 & $-$1.82 & 3.35 \\
\hline
\end{tabular}
\end{adjustbox}
\end{table}

\begin{figure}[H]
\centering

\begin{subfigure}[b]{0.32\linewidth}
    \centering
    \includegraphics[width=\linewidth]{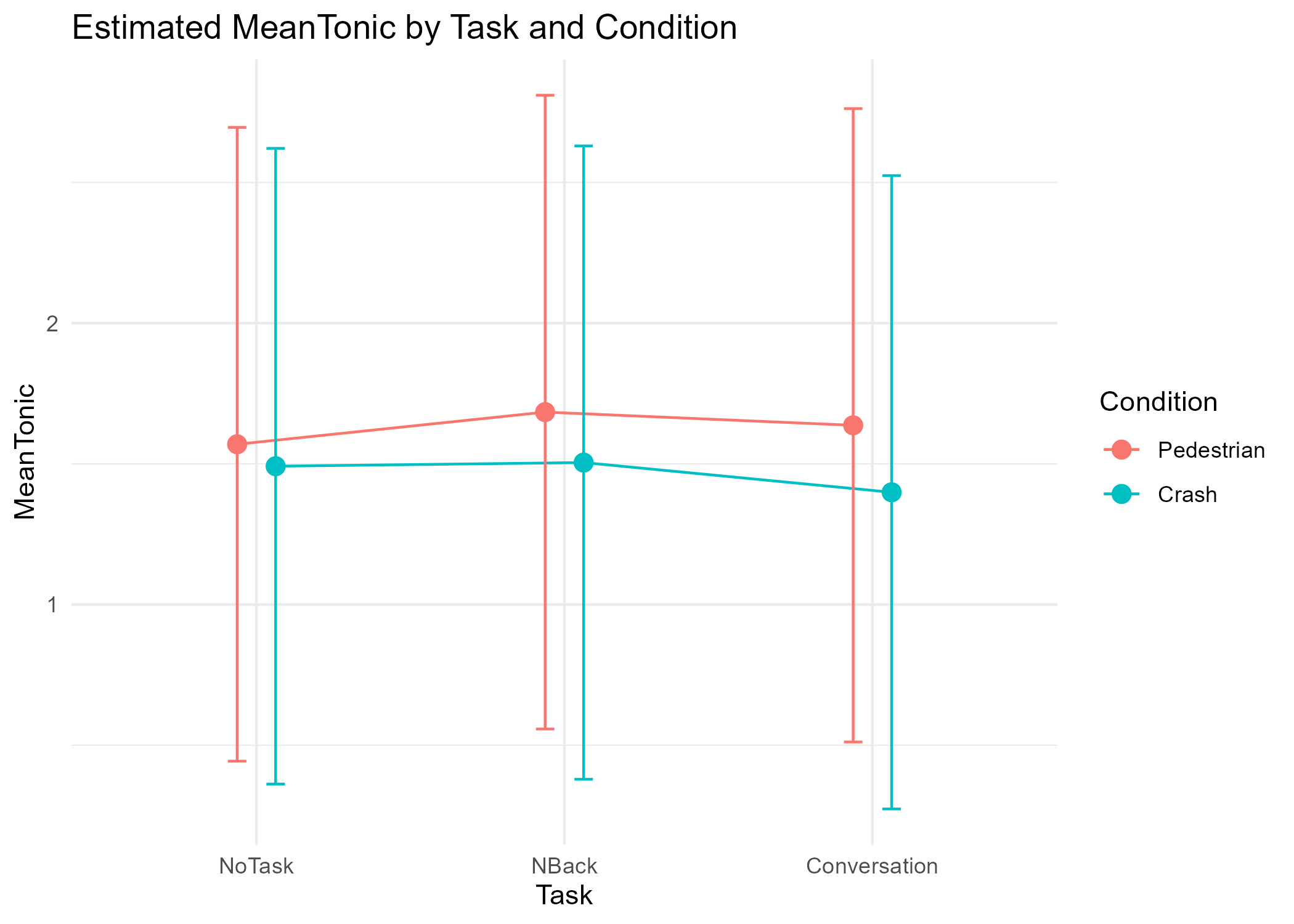}
    \caption{SCL}
    \label{fig:eda_scl}
\end{subfigure}
\hfill
\begin{subfigure}[b]{0.32\linewidth}
    \centering
    \includegraphics[width=\linewidth]{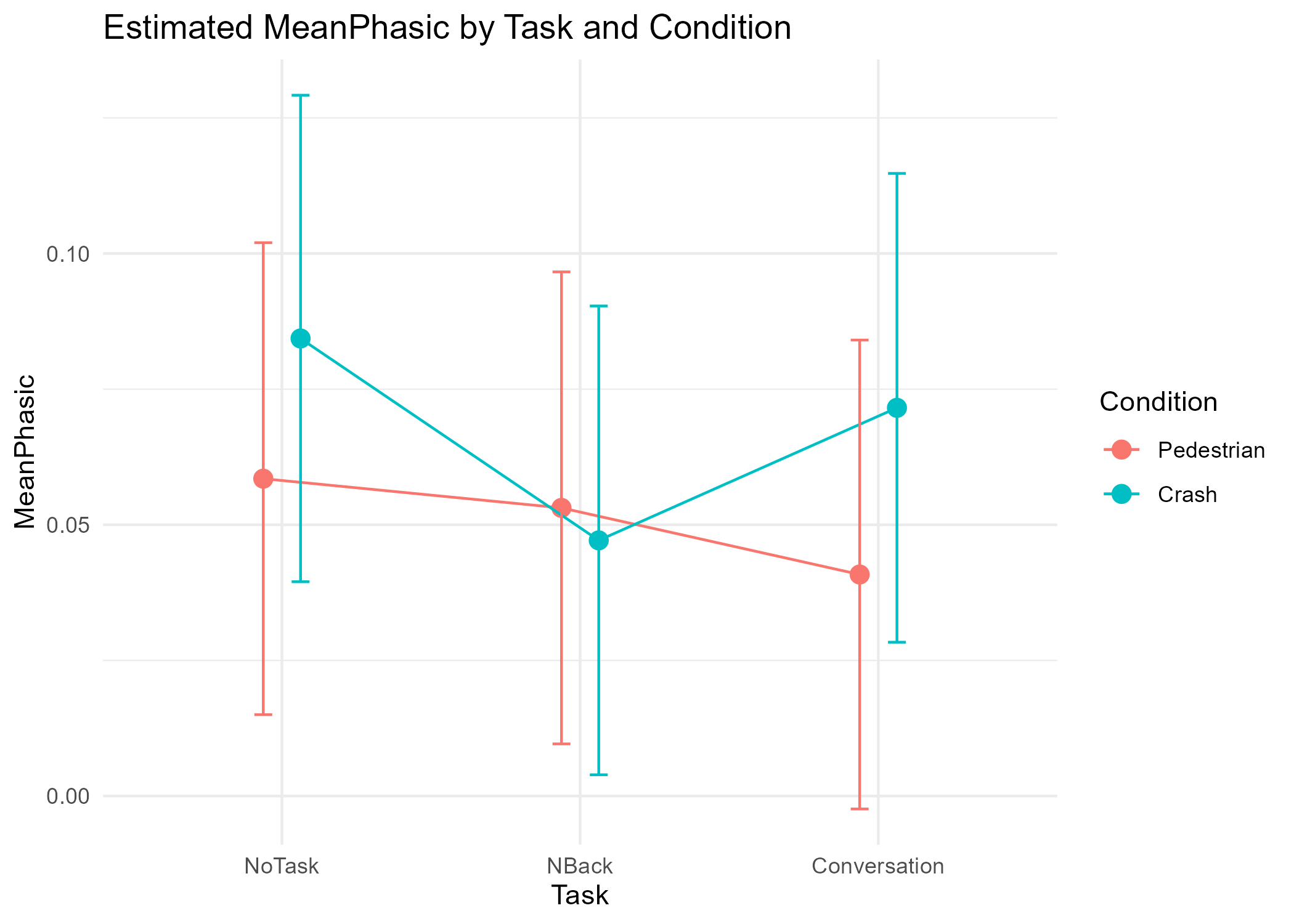}
    \caption{SCR amplitude}
    \label{fig:eda_scr_amp}
\end{subfigure}
\hfill
\begin{subfigure}[b]{0.32\linewidth}
    \centering
    \includegraphics[width=\linewidth]{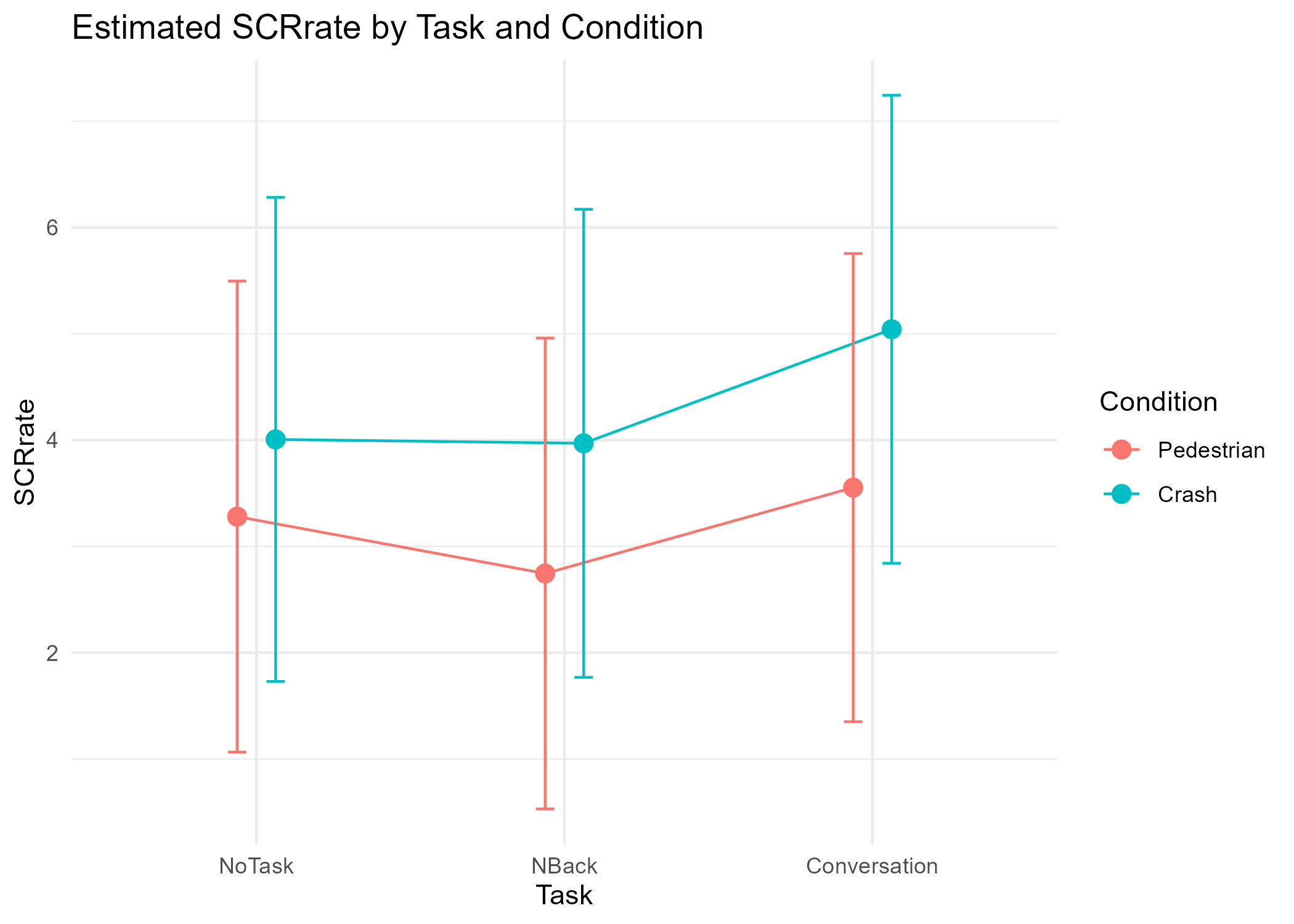}
    \caption{SCR rate}
    \label{fig:eda_scr_rate}
\end{subfigure}

\caption{Supplementary electrodermal activity (EDA) measures across task and event conditions. Error bars represent 95\% confidence intervals.}
\label{fig:eda_supplementary}

\end{figure}

\clearpage
\bibliographystyle{elsarticle-harv}
\bibliography{cas-refs}
\end{document}